\documentclass[letterpaper,twocolumn,10pt]{article}
\usepackage{usenix-2020-09}

\def\theLetterSpace{-0.5pt}
\def\extraWordSpace{-0.5pt}
\newcommand\spaceout[2][\theLetterSpace]{%
	\def\LocalLetterSpace{#1}\expandafter\spaceouthelpA#2 \relax\relax}
\def\spaceouthelpA#1 #2\relax{%
	\spaceouthelpB#1\relax\relax%
	\ifx\relax#2\else\kern\extraWordSpace\ 
	\kern\LocalLetterSpace\spaceouthelpA#2\relax\fi
}
\def\spaceouthelpB#1#2\relax{%
	#1%
	\ifx\relax#2\else
	\kern\LocalLetterSpace\spaceouthelpB#2\relax%
	\fi
}
\usepackage{microtype}

\usepackage{booktabs}

\usepackage[shortlabels]{enumitem}
\usepackage{amssymb}
\usepackage{graphicx}
\usepackage{color}

\usepackage{multirow}
\usepackage{amsmath}
\usepackage{tikz}
\usepackage{pgfplots}
\usetikzlibrary{calc}
\usepackage{pgf-umlsd}
\pgfplotsset{compat=1.16}
\usetikzlibrary{positioning}

\definecolor{editorGray}{rgb}{0.95, 0.95, 0.95}
\definecolor{editorOcher}{rgb}{1, 0.5, 0} %
\definecolor{editorGreen}{rgb}{0, 0.5, 0} %
\definecolor{backcolour}{rgb}{1, 1, 1}
\definecolor{trash}{rgb}{0, 0, 1}
\usepackage{upquote}
\usepackage{listings}

\lstdefinelanguage{js}{
	language=JavaScript,
	morekeywords={typeof, new, true, false, catch, function, return, null, catch, 
	switch, var, try, if, in, while, do, else, case, break},
	morecomment=[s]{/*}{*/},
	morecomment=[l]//,
	morestring=[b]",
	morestring=[b]',
	backgroundcolor=\color{backcolour}
}
\lstdefinelanguage{python}{
	morekeywords={typeof, new, true, false, except,
	def, return, none, try, switch, if, in, while, do, else,
	case, break, for},
	morecomment=[s]{"""}{"""},
	morecomment=[l]\#,
	morestring=[b]",
	morestring=[b]',
	backgroundcolor=\color{backcolour}
}
\lstdefinelanguage{html5}{
	language=html,
	sensitive=true, 
	alsoletter={<>=-},
	otherkeywords={
		<html>, <head>, <title>, </title>, <meta, />, </head>, <body>,
		<canvas, \/canvas>, <script>, </script>, </body>, </html>, <!, html>, <style>, </style>, ><
	},  
	ndkeywords={
		=,
		charset=, id=, width=, height=,
		border:, transform:, -moz-transform:, transition-duration:, transition-property:, 
		transition-timing-function:
	},  
	morecomment=[s]{<!--}{-->},
	tag=[s],
	backgroundcolor=\color{backcolour}
}
\makeatletter
\newcommand\footnoteref[1]{\protected@xdef\@thefnmark{\ref{#1}}\@footnotemark}
\makeatother

\definecolor{darkgreen}{cmyk}{0.4,0.1,0.4,0.4}
\definecolor{gray}{cmyk}{0.3,0.1,0.1,0.2}
\definecolor{purple}{RGB}{148, 82, 165}
\definecolor{RED}{RGB}{255, 0, 0}
\definecolor{orange}{RGB}{237, 123, 42}
\definecolor{prettydarkgreen}{RGB}{0, 101, 22}
\definecolor{greenish}{rgb}{0.078,0.62,0.48}
\definecolor{cerulean}{rgb}{0.0, 0.48, 0.65}

\newcommand{\hide}[1]{{}}

\newcommand{\todo}[1]{{\color{red}{#1}}}

\newcommand{\empirical}[1]{{\color{black}{#1}}}

\newcommand{\mi}[1]{\ensuremath{\mathit{#1}}}           %

\expandafter\def\expandafter\UrlBreaks\expandafter{\UrlBreaks
	\do\a\do\b\do\c\do\d\do\e\do\f\do\g\do\h\do\i\do\j%
	\do\k\do\l\do\m\do\n\do\o\do\p\do\q\do\r\do\s\do\t%
	\do\u\do\v\do\w\do\x\do\y\do\z\do\A\do\B\do\C\do\D%
	\do\E\do\F\do\G\do\H\do\I\do\J\do\K\do\L\do\M\do\N%
	\do\O\do\P\do\Q\do\R\do\S\do\T\do\U\do\V\do\W\do\X%
	\do\Y\do\Z
}

\definecolor{lightgray}{rgb}{.9,.9,.9}
\definecolor{darkgray}{rgb}{.4,.4,.4}
\definecolor{purple}{rgb}{0.65, 0.12, 0.82}

\lstdefinelanguage{JavaScript}{
  keywords={typeof, new, true, false, catch, function, return, null,
catch, switch, var, if, in, while, do, else, case, break},
  keywordstyle=\color{black}\bfseries,
  ndkeywords={class, export, boolean, throw, implements, import, this},
  ndkeywordstyle=\color{black}\bfseries,
  identifierstyle=\color{black},
  sensitive=false,
  comment=[l]{//},
  morecomment=[s]{/*}{*/},
  commentstyle=\color{darkgray}\ttfamily,
  stringstyle=\color{black}\ttfamily,
  morestring=[b]',
  morestring=[b]"
}

\usepackage[switch]{lineno}
\newcommand{\webdriver}[0]{{\texttt{webdriver}}}

\begin{document}
\title{
	Analysing and strengthening OpenWPM's reliability
}

\ifx\anonymise\undefined
\author{
	{\rm Benjamin Krumnow}\\
	TH K\"oln,\\Open University Netherlands
	\and
	{\rm Hugo Jonker}\\
	Open University Netherlands,\\Radboud University
	\and
	{\rm Stefan Karsch}\\
	TH K\"oln
} %
\else
\author{
	{\rm Anonymous Author(s)}
}
\fi

\maketitle
	
\begin{abstract}
Automated browsers are widely used to study the web at scale.
Their premise is that they measure what regular browsers would encounter
on the web. In practice, deviations due to detection of automation have
been found. To what extent automated browsers can be improved to reduce
such deviations has so far not been investigated in detail.
In this paper, we investigate this for a specific web automation framework:
OpenWPM, a popular research framework specifically designed to study
web privacy.
We analyse (1) detectability of OpenWPM, (2) prevalence of OpenWPM
detection, and (3) integrity of OpenWPM's data recording.

Our analysis reveals OpenWPM is easily detectable. We measure to what
extent fingerprint-based detection is already leveraged against OpenWPM
clients on 100,000 sites and observe that it is commonly detected
(\empirical{$\sim$14\%} of front pages). Moreover, we discover
integrated routines in scripts to specifically detect OpenWPM clients.
Our investigation of OpenWPM's data recording integrity identifies novel
evasion techniques and previously unknown attacks against OpenWPM's
instrumentation. 
We investigate and develop mitigations to address the identified issues. 
In conclusion, we find that reliability of automation frameworks should not
be taken for granted. Identifiability of such frameworks should be studied,
and mitigations deployed, to improve reliability.
\end{abstract}

\section{Introduction}
Web studies rely on browser automation frameworks to accrue data over
thousands of sites. The goal of such studies is to provide a view on
what regular visitors would encounter on the web. This relies on an
(often unstated) assumption that the data as collected is representative
of what a regular, human-controlled browser would encounter.
Previous works~\cite{JKV19,ADZVN20,JSSBPVLK21,CLBW22} have shown that
this is not always the case: websites have been found to omit content
(advertisements, video, JavaScript execution, login forms, etc.) or
require completion of a CAPTCHA for automated clients. While detectability
of automated browsers has been discussed online in various
blogs~\cite{She15,Vastel18,AdTechMadness} and discussion
forums,
to the best of our knowledge, so far, no in-depth academic study on
the reliability of measurement frameworks built upon such components has been performed.

In this work, we study the OpenWPM framework~\cite{EN16} for measuring
web privacy.
To date, OpenWPM has been used in at least 76 studies, 60 of which
resulted in peer-reviewed publications. As such, its fidelity, that
is, the extent to which the web OpenWPM encounters is the web as seen
by other web clients, is essential. This point is not lost on its users:
several studies explicitly remark bot detection as a possible threat for
the validity of their measurements~\cite{UDHP20,CTDSS21,LLZLDLA0HJZ0Z19}.
Recent studies investigated proliferation of generic bot 
detection~\cite{JKV19,JK19} and found over 10\% of websites employing
such techniques. However, it is not clear how these findings
translate to OpenWPM. 
On the one hand, OpenWPM uses a normal browser for collecting data,
making it harder to distinguish from other visitors. On the other,
OpenWPM targets security and privacy research, an area where malicious
actors are to be expected~\cite{SDAHN16,DMF18}. It is not clear on how
many sites OpenWPM could be detected, nor is it clear what effect such
detection may have.

In this paper, we address this point. That is, we investigate to what
extent OpenWPM provides a reliable record of how websites behave towards
any web client. First and foremost, this necessitates understanding how
OpenWPM can be distinguished from other clients. Building on that
knowledge, we provide a first estimate of how many websites are able to
distinguish OpenWPM from human visitors, finding that there are even 
several websites that can distinguish OpenWPM from other web bots. This
allows these sites to use \emph{cloaking}: responding differently to
different clients.
Secondly, we investigate whether a website can actively attack OpenWPM's
data collecting functionality. We find several new ways in which a malicious
website can attack OpenWPM's data recording.
While both cloaking and data recording attacks are possible, the question
remains whether OpenWPM-detecting sites employ such tactics. Recent work
by Cassel et al.~\cite{CLBW22} finds that Selenium-based bots receive
far less third-party traffic, which indicates cloaking happens in practice.
We develop a stealth extension for OpenWPM to prevent cloaking and test its
performance on sites where OpenWPM-detectors were found.

\ \\
\textbf{Contributions.}
Our main contributions are:
\begin{itemize}[itemsep=-1ex,topsep=0ex]
	\item \textbf{(Sec.~\ref{sec:identify_openwpm})} We provide the first analysis of 
	    OpenWPM's detectability based on both conventional
    	fingerprinting~\cite{JKV19} and template attacks~\cite{SLG19} techniques.
    	We find previously not reported, identifiable properties for every mode of
    	running OpenWPM (headless, Xvfb, etc.), even allowing to distinguish between
    	these modes. 
	
	\item \textbf{(Sec.~\ref{sec:detector_incidence})} We look for bot detectors
	    in the Tranco Top 100K sites that probe these properties 
	    via both static and dynamic analysis. We find a drastic increase of
    	Selenium-based bot detection. %
    	In addition, we find detectors in the wild specifically targeting
    	OpenWPM clients.
	
	\item \textbf{(Sec.~\ref{sec:poison_openwpm})} 
    	We explore how sites can attack OpenWPM's data collection. We find various
    	attack  vectors targeting OpenWPM's most commonly used instruments and
    	implement proof-of-concept attacks for these.
	
	\item \textbf{(Sec.~\ref{sec:hardening_openwpm})} We harden OpenWPM against
	    poisoning attacks and detection. Our hardening hides all identifiable
	    properties when run in native mode and addresses the identified attacks
	    against OpenWPM's instrumentation. We evaluate its performance against
	    vanilla OpenWPM. The number of cookies received is severely impacted.
	    Conversely, ads/tracker traffic is hardly impacted.

\end{itemize}

\begin{figure}[t]
	\centering
	\includegraphics[width=200px]{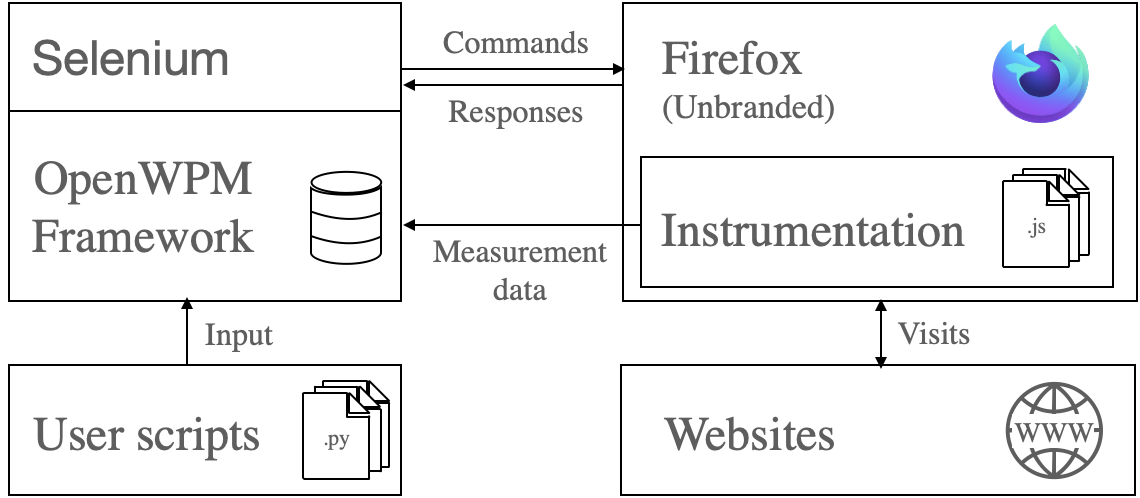}
	\caption{Components of the OpenWPM framework}
	\label{fig:openwpm_components}
\end{figure}

\section{Background}
\label{sec:background}

\paragraph{OpenWPM.}
OpenWPM is a valuable tool for web researchers, as it offers increased
stability, fidelity and easy access to measurement functionality on top
of a browser automation framework (Selenium + WebDriver).
The framework can be run under either Ubuntu or macOS.
It consists of four parts (cf.~Figure~\ref{fig:openwpm_components}):
a web client, automation components, instrumentation for measurements,
and a framework. As a web client, OpenWPM uses an unbranded Firefox
browser. In contrast to a regular Firefox browser, this allows running
unsigned browser extensions.
The various measurement instruments are
implemented as browser extensions. They facilitate recording various
website aspects, such as JavaScript calls or HTTP traffic. The last
part is the framework, which acts as the conductor. Its purpose is to 
control browsers and data collection. It also adds much-needed functionality,
such as monitoring for browser crashes and liveliness,
restoring after failures, loading input data, etc.

\paragraph{Use of OpenWPM in previous studies.}
\label{sec:literature_review}
To understand how OpenWPM is being used, we review the different studies performed
to date with OpenWPM. In December 2021, 76 works, of which 57 peer-reviewed,
were listed\footnote{\url{https://webtap.princeton.edu/software/}} as using OpenWPM. 
We further add two recent studies that had not yet been listed.
For each study, we check the following: what is measured, whether
subpages are visited, whether interaction is used, and what run mode
is used. Table~\ref{tbl:literature_review} summarises our
findings. Appendix~\ref{appdx:previous_studies} breaks down our
findings for each study individually.

The \emph{measures} category tallies how many studies used OpenWPM's
various measurement instruments: HTTP traffic, cookies, and JavaScript.
Each of these measures may be impacted individually due to bot detection.
Interestingly, while most studies use OpenWPM to record HTTP traffic, a few
(e.g.~\cite{LWPY*17,EAPN19,SIIK19,CNS20}) have used it as automation
instead as a measurement tool. These are tallied under `other' in
Table~\ref{tbl:literature_review}. 
The other categories pertain to aspects that may impact detectability. In each
case, it is currently not known whether these play a role in bot detection.
With respect to the \emph{interaction} category, we note that no study
mentioned implementing interaction mechanisms. Therefore, we assume all
studies used OpenWPM's default interaction functionality.

With respect to the \emph{run mode} category, note that not all
studies provide information about this. Nevertheless, the used run mode may 
impact detectability (e.g.~\cite{HBBP14}) and thus should be considered. 
We therefore consider all currently supported modes:
\begin{enumerate}[a.,itemsep=-3.5ex,topsep=0ex]
\item \emph{Unspecified}: study does not specify  mode, \\
\item \emph{regular}: study uses a full Firefox browsers, \\
\item \emph{headless}: study uses Firefox without a GUI, \\
\item \emph{Xvfb}: 
as regular, with visual output redirected to a buffer,   \\
\item \emph{Docker}: study runs OpenWPM within a Docker container, \\
\item \emph{Virtualisation}: study uses virtual machines, possibly in cloud 
infrastructure. 
\end{enumerate}

Lastly, we track whether the studies considered \emph{bot detection} at all
and, if so, whether they used OpenWPM's built-in anti-detection features.
Aside from studies investigating bot detection directly,
only very few consider fingerprinting~\cite{UDHP20} or 
cloaking~\cite{LLZLDLA0HJZ0Z19,CTDSS21} as a potential risk for valid
results.

\begin{table}[htb]
	\hspace{\textwidth}
	\centering
	\footnotesize
	\caption{Measurement characteristics in 60 peer-reviewed 
		studies that are built upon OpenWPM}
	\begin{tabular}{cc}
		\begin{tabular}{lr}
			\toprule
			\textbf{Category} & \textbf{Studies}\\
			\midrule
			
			\textit{Measures} & \\
			\quad -- HTTP traffic & 49 \\
			\quad -- cookies & 30 \\
			\quad -- JavaScript & 17 \\
			\quad -- other & 5 \\[1.2ex]
			
			\textit{Run mode} &\\
			\quad -- unspecified & 49\\
			\quad -- virtualisation & 14\\
			\quad -- headless &  5\\
			\quad -- regular mode & 3\\
			\quad -- Docker &  2\\
			\quad -- Xvfb  &  2
			\\[1.2ex]
			
			\bottomrule
		\end{tabular}%
		&
		\begin{tabular}{lr}
			\toprule
			\textbf{Category} & \textbf{Studies}\\
			\midrule
			
			\textit{Interaction} & \\
			\quad -- no interaction & 48\\
			\quad -- clicking & 9 \\
			\quad -- scrolling & 7 \\
			\quad -- typing & 4 \\[1.2ex]
			
			\textit{Subpages}& \\
			\quad -- not visited & 45 \\
			\quad -- visited & 15 \\[1.2ex]
			
			\textit{Bot detection} &\\
			\quad -- ignored & 46 \\
			\quad -- discussed & 14 \\
			\quad \quad $\circ$~~uses mitigation & 7 \\
			
			\bottomrule
		\end{tabular}%
	\end{tabular}%
	\label{tbl:literature_review}%
\end{table}%

\section{Related Work}
\label{sec:related_work}

\paragraph{Determining the fingerprint surface of web bots.}
Browser fingerprinting~\cite{Eckersley10} has been studied extensively in the
context of user tracking, as recently summarised in%
~\cite{LBBA20}. The idea of using fingerprinting to identify certain 
client components (such as automation frameworks) has gained more attention
recently. Vastel~\cite{Vastel18} and Shekyan~\cite{She15} conducted manual
investigations of headless browsers to pin down identifiable properties these
frameworks. %
Jonker et al.~\cite{JKV19} automated the search for identifiable properties by
using a browser fingerprinting library. They compared properties of regular 
browsers against properties of bots that belong to the same engine class. 
In contrast, Schwarz et al.~\cite{SLG19} applied a new form of fingerprinting
(JavaScript template attacks) to perform client-side vulnerability scanning. 
For a template creation, they traverse object hierarchy and store characteristics 
of each object. Later on, templates can be compared to determine the differenbce.
Finally, Vastel et al.~\cite{VRRB20} inspected bot detectors in the wild to 
collect known identifiable properties. They used these to systematically
test the responses by bot detectors to their changes.

Our work comprises the both automated approaches, by Jonker et al. and Schwarz
et al., to explore the fingerprint surface of OpenWPM. We apply these systematically 
to the various run modes of OpenWPM clients, uncovering %
distinguishers for each mode.

\paragraph{Measuring bot detection in the wild.}
Two studies exist that carried out a large-scale investigation of the
existence of unknown fingerprint-based bot detectors. Jonker et
al.~\cite{JKV19} scanned 1M websites gathering statically included
scripts and analysing these using static code analysis.
Shortly after, Jueckstock and Kapravelos~\cite{JK19}
presented a similar experiment using dynamic script collection and
dynamic analysis. Their presented tool relies on a modified V8 engine
to instrument browser functions.

\paragraph{Reliability of scraping results.}
Recently, multiple studies have been conducted that explore differences
between various automated clients, and also between automated clients
and human-driven clients in website responses.  Ahmad et
al.~\cite{ADZVN20}  investigate response differences between three
classes of bots (HTTP engine tools, headless browsers, and automated
native browsers). They found that while HTTP engine tools miss many
important resources, they more often pass bot detection than the other
two classes. Jueckstock et al.~\cite{JSSBPVLK21} studied differences
between headless Chrome and regular Chrome. For regular Chrome, they
used a puppeteer-plugin which hides distinguishable properties in Chrome
to focus on bot detection. Their results reinforce previous
recommendations~\cite{EN16,Vastel18} to not use headless browsers. Zeber
et al.~\cite{ZBORSWL20} contrast data from human users with OpenWPM
clients. In their study, OpenWPM clients encountered three times more
tracking domains and had more interaction with third-party domains than
human-controlled browsers. Cassel et al.~\cite{CLBW22} investigate the
reliability of emulated browsers. To avoid bot detection, they created
their own tooling to remotely control a browser. Interestingly, their
observations show the opposite of Zeber et al.'s findings. They observed
84\% less third party traffic for a Selenium-driven vs.~a
non-Selenium-driven Firefox browser. This contradiction shows that there
is yet no consistent picture for the influence of bot detection on
measurements. Further investigation to resolve this conundrum is needed.
Any such investigation necessitates tooling that can evade bot
detection. We aim to develop such a tool for OpenWPM.

\section{Fingerprint surface of OpenWPM}
\label{sec:identify_openwpm}
We begin by addressing the research question \textit{how can
OpenWPM be distinguished from human-controlled web clients?} In general,
a web site operator looking to identify OpenWPM clients can either 
probe for identifiable properties (i.e., fingerprinting), or attempt
to recognise OpenWPM's interaction. The latter is due to Selenium, 
whose interaction was studied in detail by Go{\ss}en et al.~\cite{GJKKR21}.
Those results fully carry over to OpenWPM. This leaves uncertainty
about how OpenWPM's fingerprint distinguishes it from other clients
and other bots. In line with previous works, we call that part
of a browser fingerprint that distinguish a certain type of client from
other types the \emph{fingerprint  surface}~\cite{TJM15}. Determining the
fingerprint surface of an OpenWPM client requires a way to find its
properties that deviate from properties and values in other clients.
Jonker et al.~\cite{JKV19} showed that it suffices to consider differences
within the client's `browser family', that is, fingerprint differences
with those clients who use the same rendering engine and JavaScript
engine. By comparing the results for multiple clients of the same
browser family, differences unique to each client are brought to light. 
In previous works, two approaches for browser fingerprinting
were used: probing a specific list of properties~\cite{JKV19}, or using an
automated approach for DOM traversion~\cite{SLG19}. While there is overlap
between the results of these methods, neither offers a complete superset
of the other. We combine the results of both approaches to determine the
fingerprint surface.

\subsection{RQ1: How recognisable is OpenWPM?}
\label{sec:fingerprint_surface}
We determine OpenWPM's fingerprint surface by comparing its client to a
standalone version of the same Firefox browser. Any differences must
originate in the hosting environment, the framework itself, the base
implementation, the added automation, or measurement components. 
To account for possible effects of the various run modes of OpenWPM 
on the fingerprint surface, we determine variations for each
setup on Ubuntu and macOS. Table~\ref{tbl:fp_surface} summarises
identifying properties found in the current version for each mode. 
In addition to ways to recognise OpenWPM's instrumentation,
we also identify ways to recognise Selenium and WebDriver,
display-less scraping (headless or Xvfb mode), and use from within a
virtual environment. Thus, every mode of running OpenWPM is identifiable
as a web bot.

\begin{table}[t]
	\centering
	\scriptsize
	\caption{Summary of deviating properties of each OpenWPM setup 
	contrasted with OpenWPM's Firefox version}
\begin{tabular}{lrrrrrr}
	\toprule
	& \multicolumn{2}{c}{\textbf{macOS}} &  
	\multicolumn{3}{c}{\textbf{Ubuntu}} & \textbf{Docker}\\
	\cmidrule(lr){2-3}\cmidrule(lr){4-6}\cmidrule(lr){7-7}
	& RM  & HM & RM & HM & Xvfb & RM\\
	\cmidrule(lr){2-3}\cmidrule(lr){4-6}\cmidrule(lr){7-7}
	navigator.webdriver is true & \checkmark  & \checkmark & \checkmark & 
	\checkmark & \checkmark &  \checkmark\\
	screen dimension prop. & \checkmark  & \checkmark &
	\checkmark & \checkmark& \checkmark &  \checkmark\\
	screen position prop. & \checkmark  & \checkmark  &
	\checkmark & \checkmark& \checkmark & \checkmark  \\
	font enumeration & -- & -- & -- & -- & -- &  \checkmark  \\
	timezone is 0 & -- & -- & -- & -- & -- & \checkmark \\
	\cmidrule(lr){2-3}\cmidrule(lr){4-7}
	navigator.languages prop. & -- &   43 & -- &   43 & -- & -- \\
	deviating WebGL prop.     & -- & 2037 & -- & 2061 & 18 &  27 \\[1.2ex]
	\emph{With instrumentation:} \\
	- through tampering  & +253  & +253 & +252 & +252 & +252 & +252  \\ 
	- added custom functions & +1 & +1  & +1  & +1 & +1 & +1 \\ 
	\bottomrule
\end{tabular}%
\ \\
\vspace{0.2cm}
\footnotesize
RM: Regular mode;
HM: Headless mode;
Xvfb: X virtual frame buffer mode. \\
\label{tbl:fp_surface}
\end{table}

\paragraph{Recognising automation components.} 
We found three identification measures that may be applied against 
all modes of OpenWPM. First, the \texttt{navigator.webdriver} property
indicates a browser controlled via the WebDriver 
interface.\footnote{\url{https://www.w3.org/TR/webdriver2/\#example-1}}
Second, screen properties use standard values and cannot be changed
from OpenWPM (see Table~\ref{tbl:screen_properties_in_bots}).
For example, the browser window position for macOS is 4px from the top
and 23px from the left. On macOS, all browser instances will use the
same absolute coordinates, while on Ubuntu, each window is shifted by
the same offset, when using regular mode.

\begin{table}[h]
    \centering
    \scriptsize
    \caption{Screen properties for various configurations}
    \begin{tabular}{llr@{ x }rr@{ x }rrrr}
    \toprule
    \textbf{OS} & \textbf{Mode} & \multicolumn{2}{c}{\textbf{Resolution}} & 
    \multicolumn{2}{c}{\textbf{Window}} & \textbf{X} & \textbf{Y} & \textbf{Offset 
    (x, 
    y)}\\
    \midrule
    macOS   & Regular   & 2560 & 1440 & 1366 &  683 & 23 & 4 & 0, 0\\
            & Headless  & 1366 &  768 & 1366 &  683 & 4 & 4 & 0, 0\\
    Ubuntu  & Regular   & 2560 & 1440 & 1366 &  683 &  80 & 35 & 
    8, 8\\
            & Headless  & 1366 &  768 & 1366 &  683 &  0 & 0 & 0, 0\\
            & Xvfb      & 1366 &  768 & 1366 &  683 &  0 & 0 & 0, 0\\
            & Docker    & 2560 & 1440 & 1366 & 683 &  0 & 0 & 0, 0\\
    \bottomrule
    \end{tabular}
    \label{tbl:screen_properties_in_bots}
\end{table}

\paragraph{Identification of missing displays.}
Suppressing output to display (by using Xvfb, headless, or
Docker) adds a significant number of differences. In headless mode, the
lack of a WebGL implementation leads to thousands of missing properties.
We also observe that this mode adds 43 new properties to the
\texttt{navigator.language} object. Xvfb mode uses a regular Firefox
browser, which contains WebGL functionality. Nevertheless, Xvfb mode
causes \empirical{5} changed and \empirical{13} missing properties.
Interestingly, both headless and Xvfb mode allow the detection of
missing user elements by accessing the property \emph{screen.availTop}.
This describes the first y-coordinate that does not belong to the user
interface\footnote{\url{https://developer.mozilla.org/en-US/docs/Web/API/Screen/availTop}}.
 In display-less modes, this is always zero, while regular browsers have larger 
values.

\paragraph{Traces of virtualisation.}
Using OpenWPM's docker container causes the WebGL vendor property to contain the term
\texttt{VMware, Inc.} (cf.~Table~\ref{tbl:webgl_display-less_modes}) --
clear evidence for the use of virtualisation. In addition, the Docker environment
reduces the number of available JavaScript fonts to one (Bitstream Vera Sans Mono),
nor does it provide information about the time zone.

\begin{table}[bh]
\centering
\footnotesize
\caption{Selected deviations in display-less modes on Ubuntu}
\begin{tabular}{llr}
	\toprule
	\textbf{Mode} & \textbf{WebGL vendors} & \textbf{avail\{Top|Left\}}\\
	\midrule
	RM          & \emph{AMD~AMD TAHITI} & 27, 72\\
	HM          & \emph{Null} & 0, 0\\
	Xvfb        & \emph{Mesa/X.org$~$llvmpipe (LLVM 12.0.0,\ldots)} & 0, 0\\
	Docker      & \emph{VMware, Inc.$~$llvmpipe (LLVM 10.0.0,\ldots)} & 27, 72 
	\\
	\bottomrule
\end{tabular}
\label{tbl:webgl_display-less_modes}
\end{table}

\begin{figure}[t]
	\centering
	\includegraphics[width=200px]{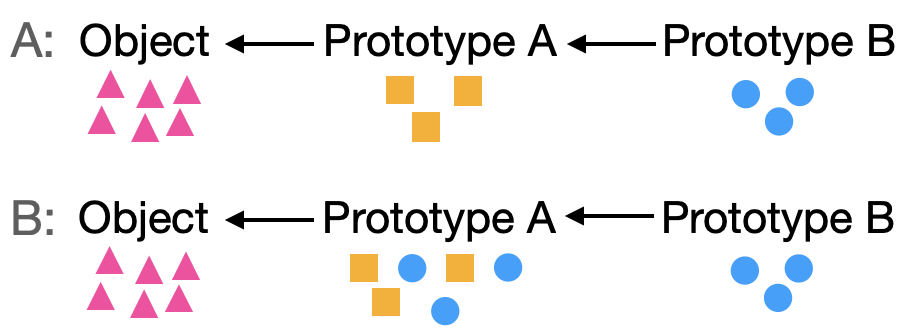}
	\caption{Properties in a (A) original object or 
	(B) by the instrumentation polluted object.}
	\label{fig:pollution}
\end{figure}

\paragraph{Detecting instrumentation.}
\label{sec:detecting_instrumentation}
We checked if using any of OpenWPM's various instruments has any effect
on its fingerprint surface. The only differences occur when using the
JavaScript instrument. First, this instrument overwrites certain of the
browser's standard JavaScript objects, which can be detected by using
the \texttt{toString} function of a function or object (see
Listing~\ref{lst:openwpm_javascript}). Another identifying aspect of
this instrument is the presence of a function in the window object
(\texttt{window.getInstrumentJS}), which is not present in any common
desktop browser (Firefox, Safari, Chrome, Edge, Opera). Third, OpenWPM's
wrapper functions can be found in stack traces. For that, a script need
to provoke an error in any overwritten function and catch the stack trace
to successfully identify a modification by OpenWPM. Lastly, the instrument
`pollutes' prototypes along the prototype chain of an object. Instrumenting
is done by changing the prototype of an object, as well as all its ancestor
prototypes. However, the properties of later ancestor prototypes are all
added to the first ancestor prototype (cf., Fig.~\ref{fig:pollution}).
This distinguishes a visitor with  instrumentation from one without.

\begin{lstlisting}[frame=tb, caption=Detectability of OpenWPM's JavaScript instrumentation, 
	language=js,label=lst:openwpm_javascript, columns=fullflexible,
	basicstyle=\scriptsize,float=bp]
	window.canvas.getContext.toString();
	
	// output of .toString when not instrumented
	"function getContext() { 
		[native code] 
	}"
	
	// output of .toString when instrumented
	"function () { 
		const callContext = \
		getOriginatingScriptContext(!!logSettings.logCallStack);
		logCall(objectName + "." + \
		methodName, arguments, callContext, logSettings); 
		return func.apply(this, arguments); 
	}"
\end{lstlisting}

\paragraph{Evaluation.} %
\label{sec:fingerprint_based_detector}
We validate whether the identified fingerprint surface works in
practice to identify OpenWPM. For that, we implemented a OpenWPM
detector, that uses four tests to identify OpenWPM amongst web
clients: (1) test for the presence of a DOM property, (2) test for a
missing DOM property, (3) test if a native function was overwritten,
(4) compare a DOM property with an expected value.

We tested the detector by setting up four machines, 2 Macintoshes
and 2 PCs with Ubuntu. On each machine, we used OpenWPM and common
browsers (Chrome, Safari, Opera and Firefox). We tested each
distinguishing property from Table~\ref{tbl:fp_surface}. Our
detector site was able to correctly identify OpenWPM every single
time. Almost all properties uniquely identify OpenWPM, except for a
few WebGL- and screen-related properties. For a few WebGL properties
(roughly 200 of 4K), we found that these also occur on some
non-OpenWPM clients. Ignoring all such properties still leaves a
large number of identifying properties.

\subsection{RQ2: How stable is the fingerprint surface?}
\label{sec:fingerprint_stability}
We explored how stable our determined fingerprint surface is, as new
Firefox and OpenWPM versions may appear frequently. To that end, we
repeated our experiments for older version of OpenWPM (0.11.0 and 0.10.0).
In general, we found that the fingerprint surfaces largely overlap.
For example, on MacOS, the number of WebGL deviations in headless mode
increases to 2037 in OpenWPM 0.17.0, from OpenWPM 0.11.0's 2022.
In the oldest OpenWPM version (0.10.0), we find that the JavaScript
instrument adds two properties instead of one to the window object
(\texttt{jsInstruments} and \texttt{instrumentFingerprintingApis}).
In addition, we also investigated whether using an unbranded
browser (as OpenWPM does) impacts OpenWPM's fingerprint.
We found no differences between branded and unbranded Firefox versions.

Using outdated browsers, however, does impact the fingerprint.
For example, Google's reCAPTCHA service assigns a higher risk to
older browser variants~\cite{SPK16_Blackhat}. In the past, OpenWPM's
integrated Firefox version has been behind the official release
of Firefox several times (cf.~Table~\ref{tbl:firefox_releases} in
Appendix~\ref{appdx:fx_versions}). We found that OpenWPM used an
outdated Firefox browser \empirical{71\%} of the last 20 months. In short,
this distinction vector should be expected when using OpenWPM.

\section{Incidence of OpenWPM detection}
\label{sec:detector_incidence}
To assess the extent of OpenWPM detection in the wild, we conduct a large-scale 
measurement for client-side bot detection. In detail, we focus on scripts with 
capabilities to detect OpenWPM, i.e. scripts with routines to access properties 
unique for Selenium-based bots and/or OpenWPM. 
We find both general Selenium detectors and OpenWPM-specific detectors.

\subsection{Data acquisition and classification}
\paragraph{Methodology.}
Previous automated approaches~\cite{JKV19,JK19} to identify bot
detectors have either relied on static or dynamic analysis. The idea
behind static analysis is to identify code patterns in source code that
link to known bot detectors or that use specific bot-related properties.
A limitation is that scripts may create code dynamically, which will be
missed out by static analysis. Moreover, minification and obfuscation
further increase the false negative rate of static analysis.
The alternative approach, dynamic analysis, is to monitor JavaScript
calls that identify a script as bot detector based on access to 
bot-related properties. Dynamic analysis does cover dynamically-generated
scripts. Moreover, it does not monitor the code itself, but only executed
calls. An upside of this is that neither minification nor obfuscation
affects the analysis. On the other hand, code that happens not to be
executed during the run, is not analysed. Both static and dynamic analysis 
have been able to identify some bot detectors in the wild. 
It is not clear whether and to what extent the results of the methods differ
in practice for finding web bot detectors.
We combine both methods to increase coverage.

\paragraph{Setup.}
In order to assess the extent of client-side bot detection, we scan the
top 100K websites of the Tranco 
list~\cite{TrancoList}\footnote{\url{https://tranco-list.eu/list/WV79}}.
We set up an instance of OpenWPM running Firefox in regular mode. During a site
visit, our OpenWPM client stores a copy of any transmitted JavaScript
file and records JavaScript calls. We add an initial waiting time of 45
seconds after a completed page load to give websites enough time to
perform JavaScript operations. In addition, we instruct our
client measures the presence of bot detection on subpages by opening a
maximum of three URLs extracted from a site's landing page. For
selecting subpages, we consider only URLs linking to the same domain.
Within this and the following sections, we apply scheme \texttt{eTLD+1}
to identify a domain. To account for websites that use same origin
requests to redirecting clients to foreign domains, our client checks if
a foreign domain was entered after following all redirects.

Scripts should be classified as bot detectors if they access the fingerprint
surface of OpenWPM.
However, certain scripts may access these attributes for other purposes, such
as checking supported WebGL functionality. To reduce such false positives,
we only classify a script as bot-detecting when it accesses properties
pertaining to browser automation or are unique to OpenWPM 
(cf.~Sec.~\ref{sec:fingerprint_surface}). This leaves only the following:
\texttt{navigator.webdriver}, which is specific
to WebDriver-controlled bots; and the new identifying properties introduced by OpenWPM's JavaScript
instrumentation:
\texttt{getInstrumentJS},
\texttt{instrumentFingerprintingApis}, and
\texttt{jsInstruments}.
Table~\ref{tbl:prevalence-of-detectors} shows the results of the data
collection and classification.

\paragraph{Limitations.} 
Inherent in the above approach are several assumptions that can impact
the results. First, our approach relies on the fingerprint surface we established.
Detectors based on other methods (e.g., mouse
tracking~\cite{CGKWJ13}) will be missed. Second, we do not
account for cross-site tracking. A third-party tracker could classify 
our client as a bot on one site and would need only to re-identify the client on
another site, e.g., using IP filtering or regular browser fingerprinting.
This amounts to a form of \emph{website cloaking} -- serving different content
to specific clients. To what extent third party tracking in general employs
cloaking is a different study and left to future work.
Both these limitations may cause underestimation of the number of
detectors (false negatives). As such, our approach approximates a
lower bound on the number of detectors in the wild.

\paragraph{Preprocessing for static analysis.}
Within the static analysis, we pre-process scripts to undo
straightforward obfuscation. We derive the respective encoding,
transform hex literals to ASCII characters, and remove code comments.
We apply our static analysis to scripts that we collected during our scan 
of the Tranco Top 100K, which resulted in \empirical{1,535,306} unique scripts.
To identify Selenium-detector scripts, we then use patterns to look for
access to \texttt{navigator.webdriver} (more details can be found in Appendix~\ref{appdx:patterns}).

\paragraph{Using honey properties to catch iterators.}
For the dynamic analysis, every recorded access to the fingerprint
surface identifies a script with the potential to detect OpenWPM as a
bot. This will also be triggered by scripts that iterate over all
properties, e.g., for regular browser fingerprinting (re-identification). 
Determining the purpose of such iteration requires per-script manual
inspection and goes beyond dynamic analysis.

To determine whether property iteration takes place, we extend our
client's navigator and window object with `honey' properties. These
honey properties are added on the fly and use random strings as name.
Hence, only a script using property iteration would access all honey
properties. We assign scripts
that use property iteration into three categories, based on access to
the \texttt{navigator.}\webdriver{} property: definitely detecting bots,
and inconclusive. 
Iterator scripts are classified as inconclusive if they do not
access \texttt{navigator.}\webdriver{}, as all accesses to the fingerprint
surface could be due to property iteration.
Scripts that iterate the navigator object will naturally access the
\webdriver{} property. To check whether this access is only by iteration
or intentional, we distinguish between scripts that trigger our static
analysis and those that do not. Only scripts that do not surface in the static
analysis are classified as inconclusive. 

\subsection{RQ3: How often is OpenWPM detected?}
\label{sec:wild_detector_results}
\begin{table}
\centering
\footnotesize
\caption{Number of websites with Selenium detectors}
\begin{tabular}{lrrr}
\toprule
\textbf{\# sites}   & \textbf{static} & \textbf{dynamic} & \textbf{union}\\
\midrule
identified & 32,694 & 19,139 & 38,264\\
without false positives / `inconclusive' & 15,838 & 16,762 &  18,714\\

\bottomrule
\end{tabular}
\label{tbl:prevalence-of-detectors}
\end{table}

Our results show that, when checking both front- and
subpages, at least \empirical{16.7\%} of websites in the
Tranco Top 100K execute scripts that accessed properties specific to
Selenium and, thereby, OpenWPM. Moreover, we also find scripts
accessing OpenWPM-specific properties.

\paragraph{OpenWPM-specific properties are accessed in the wild.}
Most scripts we found recognise OpenWPM by targeting Selenium.
A small number of detectors, also include specific routines to
detect OpenWPM itself. Overall, \empirical{356} sites executed
scripts that accessed OpenWPM-specific properties. These scripts
were all included via third-party domains, belonging to four distinct providers.
Table~\ref{tbl:specific_openwpm_detectors} summarises these detectors
and their detection method. Detectors on
cheqzone.com were found by both static and dynamic analysis; 
detectors on the other three domains used some form of
minification, obfuscation, and/or dynamic loading, and were only found
by dynamic analysis. 
We investigated the four hosting domains by consulting \texttt{whois}
records, EasyList,\footnote{\url{https://easylist.to/easylist/easylist.txt}}
and the WhoTracksMe database~\cite{WhoTracksMe}.
All domains are related to the advertising industry. The domain 
\texttt{cheqzone.com} belongs to CHEQ, %
a company fighting ad fraud. The scripts hosted by Google domains are included
through Google's reCAPTCHA service.
While we could not clarify the origin of \texttt{adzouk1tag.com}, we found 
this domain listed in the EasyList for ad domains.

\begin{table}
\centering
\footnotesize
\caption{Number of sites ordered by script domains accessing
OpenWPM-specific properties}
\begin{tabular}{lrrrr}
\toprule
&  \textbf{cz} & \textbf{gs} & \textbf{google.com} & \textbf{ad1t} \\
\midrule
\textbf{total} &\textbf{331} & \textbf{14} & \textbf{9} & \textbf{2}\\
jsInstruments&331&5& 2 &2\\
instrumentFingerprintingApis & 0 &6 & 4 & 0\\
getInstrumentJS & 0 & 3 & 3 & 0\\
\bottomrule
\end{tabular}
\label{tbl:specific_openwpm_detectors}
\ \\[1.5ex]
\begin{minipage}{.5\textwidth}
	\centering
	cz:~cheqzone.com, gs:~googlesyndication.com, ad1t:~adzouk1tag.com
\end{minipage}
\end{table}

\begin{figure}
\centering
\includegraphics[width=220px]{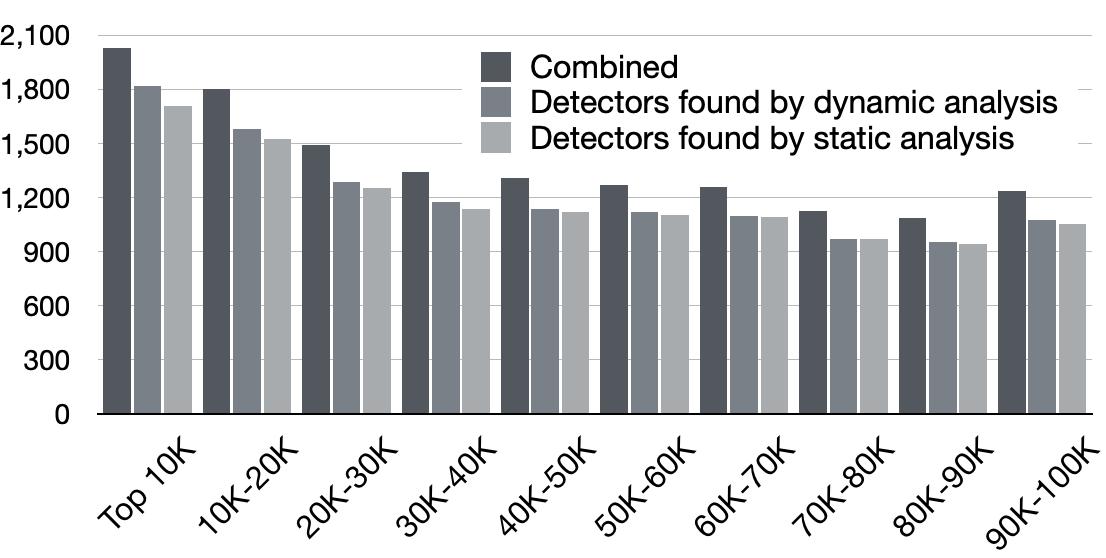}
\caption{Detectors found on front pages}
\label{fig:distribution_detectors}
\end{figure}

\paragraph{14\% of sites have bot detection on the front page.}
Figure~\ref{fig:distribution_detectors} depicts the distribution for
detectors active on the front page of websites for static and dynamic
analysis. Dynamic analysis without considering property iteration
identifies \empirical{12,208} sites with detectors on the front page.
Static analysis measures the number of sites where bot detection
could be triggered (\empirical{11,897}), including those where detection
is present but not (yet) executed, e.g., where detection is only
triggered after hovering over certain elements. While both static and
dynamic analysis identify a similar number of detectors for each bucket,
they do not fully overlap. Combining both provides a slight
increase in the presence of detectors (\empirical{$\sim$1.7K} sites).

\begin{figure}
\centering
\includegraphics[width=240px]{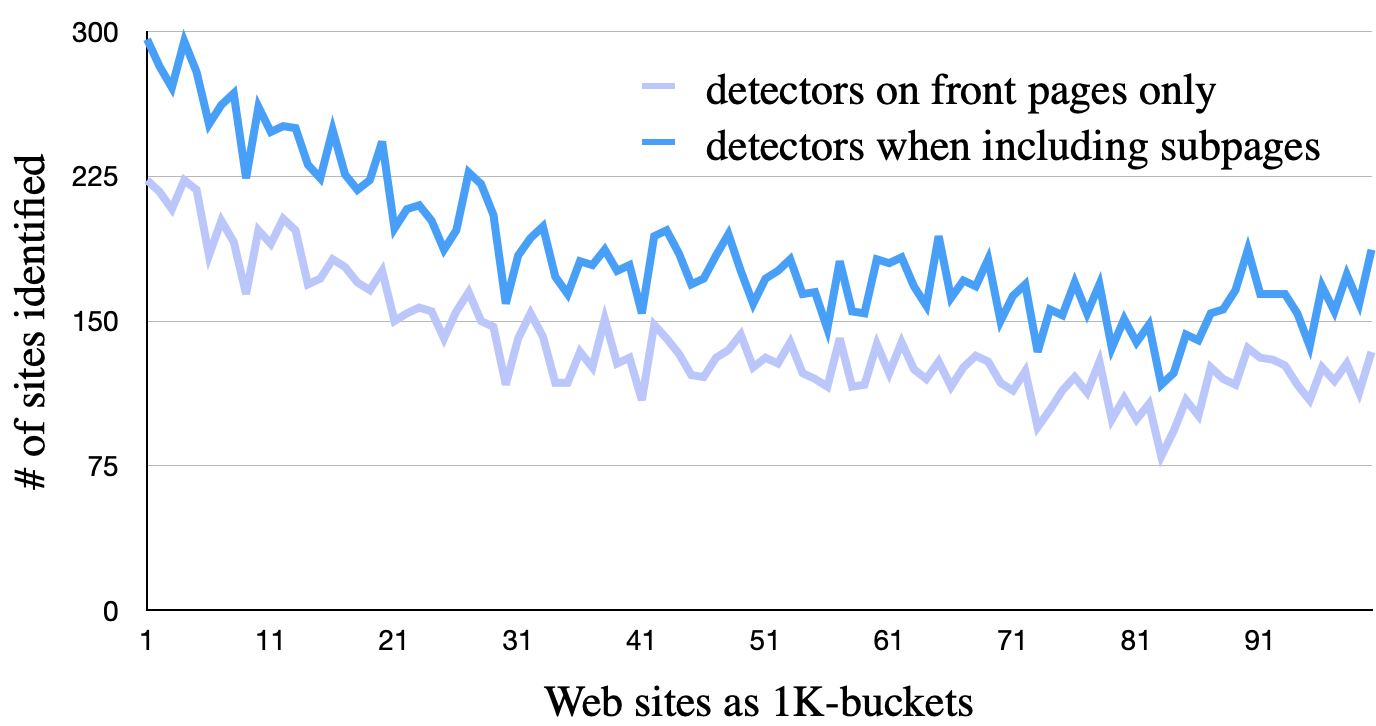}
\caption{Number of sites with bot detectors on front- and subpages
	(depicted per 1K sites)
}
\label{fig:bot_detectors_subpages}
\end{figure}

\paragraph{Deep scanning increases the rate of detection by 5 per
cent points.}
As discussed in Section~\ref{sec:literature_review}, 15\%
of studies conducted with OpenWPM (also) investigated subpages. This
raises the question whether such studies are more often subject to bot
detection, that is: does bot detection occur more frequently on
subpages? Figure~\ref{fig:bot_detectors_subpages} depicts the occurrence
of bot detectors on front pages and subpages. In general, studies
examining subpages are at greater risk to be detected: the number of
sites with active detectors increases for by at least \empirical{37}\%.
Hence, the average detection rate within the Top 100K sites will
increase. That is: the study will be exposed to more detectors.
Combining the results of both measurements, we see an increase of
\empirical{5} per cent points (from \empirical{14}\% to
\empirical{19}\%).

\subsection{RQ4: By whom is OpenWPM detected?}
\label{sec:detectors_analysis}
To explore this question, we separated detectors into first and
third parties. We find that the majority of sites includes detectors
from third-party domains. We count how often scripts on these third-party
domains are included on scanned sites, tallying each third-party domain
once per including site. Some sites include more than one detector,
hence the total number of inclusions exceeds the number of sites with
detectors. Overall, we count \empirical{3,867} first-party detector
scripts and \empirical{21,325} third-party detector scripts. 

\begin{figure}
\centering
\includegraphics[width=230px]{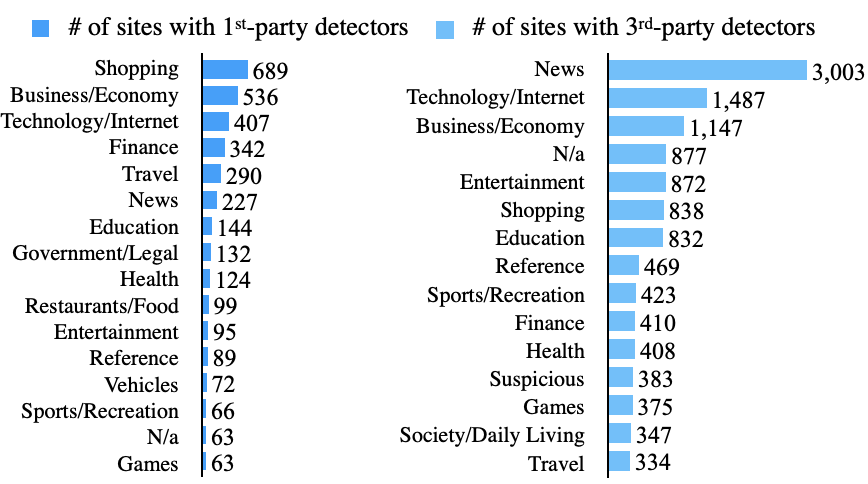}
\caption{Common categories of sites including detectors}
\label{fig:website_categories}
\end{figure}

\paragraph{First and third-party bot detection are used differently
among the industry.}
\label{sec:overview_top_third_party_detectors}
We further explore what sites include detectors, as this may provide a
better view on what bot detection is used for. For that, we collect
categories for the identified 16K websites with detectors based on
Symantec's site review service (\url{https://sitereview.norton.com/}).
Sites may be assigned multiple categories; for such sites, we tally each
listed category. Figure~\ref{fig:website_categories} depicts the 16 most
often tallied categories for both first-party detectors
(\empirical{4,198} times) and third-party detectors (\empirical{16,323}
times). We find that news sites are responsible for \empirical{18.4}\%
of all third-party inclusions, followed by Technology (\empirical{9\%}) and 
Business
(\empirical{7\%}). Interestingly, the ranks for Shopping (\empirical{16.4\%}) 
and News (\empirical{5\%})
switch for first-party detector inclusions. Moreover, sites in the
categories Finance (\empirical{8\%} vs \empirical{3\%}) and Travel  
(\empirical{7\%} vs \empirical{2\%}) make up for a
larger portion in the set of first-party inclusions than for third
parties.

We believe that these uneven distribution of inclusions is explainable.
While every site owner will want to protect their site from nefarious bots
(and thus reason to include first-party detection),
advertising has become a popular business model for websites. For such
sites, third parties have a vested interest in detecting bots:
to detect ad fraud. Thus, on such sites, one would expect more third-party
bot detection.

\paragraph{Third-party bot detection typically serves the 
advertisement industry.}
Following up on the previous point, we investigated the origins of third
party detectors.
Table~\ref{tbl:most_common_detectors} breaks down the most common
included domains. %
The top 10 domains account for two third of inclusions. The site
WhoTracks.me~\cite{WhoTracksMe} categorises trackers according to
purpose. Using this, we find that the bot-detecting scripts on the most
commonly included domains can serve a variety of purposes. For example,
\text{yandex.ru} offers scripts used for advertising, content delivery
network, site analytics, social media, and others. Other uses include
web analytics (crazyegg.com), CDN (jsdelivr.net) and live chat
(intercomcdn.com). However, bot detection is most commonly deployed by
advertisers (e.g., domains 2,3,4,7,9, and 10 in
Table~\ref{tbl:most_common_detectors}).

\begin{table}
\centering
\footnotesize
\caption{Domains hosting 3$^{rd}$-party detector scripts}
\begin{tabular}{rlrr}
	\toprule
	& \textbf{hosting domain} & \# \textbf{inclusions (1/site)} &
	\textbf{\%}\\
	\midrule
	& \textit{all} & \textit{21,325} & \textit{100\%}\\
	1 & yandex.ru & 3,848 & 18.04\% \\
	2 & adsafeprotected.com & 2,309 & 10.83\% \\
	3 & moatads.com & 2,165 & 10.15\% \\
	4 & webgains.io & 2,091 & 9.81\% \\
	5 & crazyegg.com & 1,552 & 7.28\% \\
	6 & intercomcdn.com & 1,061 & 4.98\% \\
	7 & teads.tv & 854 & 4.00\% \\
	8 & jsdelivr.net & 423 & 1.98\% \\
	9 & mxcdn.net & 416 & 1.95\% \\
	10 & mgid.com & 402 & 1.89\% \\
	11+& \emph{remaining 704 domains} & \emph{6,204} & 
	\emph{29.1\%}\\
	\bottomrule
\end{tabular}%
\label{tbl:most_common_detectors}%
\end{table}%

\paragraph{The vast majority of first-party detectors are embedded third 
parties.}
To determine the origins of first-party bot detection scripts, we look
for similarities between their inclusions of detectors. To do so, we
hash the scripts and check for structural similarities in script URLs
(for more details see Appendix~\ref{appdx:first_party_detector_details}).
We found various similarities amongst unrelated sites. Scripts
originating from Akamai occur the most frequent (\empirical{1,004}
sites). Second is Incapsula (\empirical{998} sites), third is an unknown
bot detector (\empirical{659} sites), and fourth is Cloudflare
(\empirical{486} sites). Together, these top three originators account
for \empirical{3,147} out of \empirical{3,867} sites (\empirical{88\%})
where we found first-party detectors.
In contrast to the purpose of third-party detectors, first-party
detectors are not supplied by advertisement companies. Moreover, Akamai,
Incapsula and Cloudflare all offer commercial bot detection services.
With that in mind, one should expect sites with first-party detectors to
likely tailor their responses for detected bots (e.g.,  throttling,
blocking, withholding resources, and serving CAPTCHAs).

\section{Attacking JavaScript recording}
\label{sec:poison_openwpm}

We investigate whether a malicious  website or third party could corrupt
OpenWPM's data collection process. In particular, we consider an attacker
that can deliver arbitrary content (HTML, cookies, JavaScript), but cannot
break the browser's security model. To do so, our focus resides on attacks
against the integrity or completeness of measurements. More specifically,
we aim to attack the resilience of OpenWPM's most commonly used instruments:
HTTP traffic, cookie recording, and JavaScript call recording.
Both HTTP and cookie instruments are simple wrappers around browser functionality.
Breaking them thus requires breaking the browser, which is outside the attacker
model. The JavaScript instrument, on the other hand, needs to supply all its
monitoring functionality itself.  It is therefore clearly in scope of
our attacker model.

Since the instruments focus on data recording, we investigate attacks on data
recording. More specifically, we consider:
\begin{enumerate}[1.,itemsep=-1ex,topsep=0ex]
\item whether data recording may be prevented;
\item whether fake data can be injected into the data recorder;
\item whether already recorded data can be deleted or altered;
\item finally, whether the data recording is complete.
\end{enumerate}

Instruments in OpenWPM are implemented as a browser extension.
Extensions are isolated to protect higher privilege APIs from access
by untrusted code. Website scripts thus cannot directly interact with
extensions. However, both extensions and website scripts can read and
change the DOM, opening the door for injection attacks against extensions
that read the DOM.
We conducted source code analysis for each instrument under investigation
to identify vulnerabilities to such attacks. Below we discuss the found
vulnerabilities.

\subsection{RQ5: How to prevent data recording?}
\label{sub:turn_off_recording}
We found a vulnerability that, when successfully
exploited, allows a website to break OpenWPM's data recording hooks. 
The vulnerability can be leveraged to turn off
recording of JavaScript calls in the JavaScript instrument. 

More specifically: the JavaScript instrument overwrites several API functions
by hooking into the DOM's event dispatcher (to record access to them). The event
dispatcher then sends messages to be recorded back to the JavaScript instrument's
back end. To prevent an attacker from silently undoing these hooks, OpenWPM
also hooks into (and thus: records access to) setters and getters to these API
functions themselves. However, the event dispatcher itself is not protected.
We thus can alter the event dispatcher to inject our
own messages and manipulate messages sent to OpenWPM 
(cf., Listing.~\ref{lst:scrip_recorder}). To carry out this attack, the attacker
overrides the event dispatcher to block all messages (all events from instrumented
objects). This would already block OpenWPM recording, by breaking any JavaScript
API calls. However, this also would break a website's own JavaScript. 
To block only OpenWPM messages, the block needs to be tailored. Conveniently,
tags messages with an ID to identify any monitored objects. Though this ID is
randomly generated, it can easily be determined: simply trigger an API
call to a monitored object, acquire the random ID from the observed message, and
update the event dispatcher to only block messages containing this ID.

\begin{lstlisting}[frame=tb, caption=Turn off the script recorder, 
    language=js,label=lst:scrip_recorder, columns=fullflexible,
    basicstyle=\scriptsize,float=tbh]	
//Step I: Retrieve OpenWPM's random ID
function grabID() { return new Promise((resolve, reject) => {
  let id;
  document.dispatchEvent = function (event) {
    id = event.type; document.dispatchEvent = dispatch_fn;
    if (id !== undefined) { resolve(id);
    } else { reject(new Error(msg));}
  }
  // Perform an action to grab the ID
  navigator.userAgent;});}
// Step II: Overwrite event dispatcher to block events
async function attackExtension() {
  let id = await grabID();
  document.dispatchEvent = (event) => {
    if (event.type != id) { dispatch_fn(event); // Dispatch event
    } else {console.log("Event swallowed: " + event);}}}
\end{lstlisting}

\subsection{RQ6: Can fake data be injected?}
\label{sec:fake-data-injection}
The previous attack, altering the event dispatcher, not only allows an
attacker to block data recording, it also allows an attacker to learn
the ID OpenWPM uses to record data. This is sufficient to inject almost
arbitrary messages to be recorded. The attacker simply creates a custom
event following the format used by OpenWPM's JavaScript extension and
includes OpenWPM's assigned event ID. This enables an attacker to define
most of the content of the resulting entry in OpenWPM's recording, such
as the executing script URL or which function was called. Crucially,
though, the website that originated the call is set outside of the
browser by OpenWPM. The data sent by the event dispatcher is properly
sanitized by the back-end, which prevents spoofing this. We can thus
only inject fake data for the currently visited website. Note that a
third party included on the site can also execute this attack.

\subsection{RQ7: Can records be deleted or altered?}
\label{sec:alter-delete-data}
Whereas the previous attacks exploited a vulnerability in the DOM-parsing
front-end of the respective instruments, deleting already recorded data
requires manipulating an instrument's back-end, for OpenWPM: SQLite.
Attacking a database back-end requires an SQL injection vulnerability.
As already mentioned, OpenWPM's data recording back-end properly sanitizes
its inputs. This means that there is no possibility for an SQL injection via
JavaScript recording. Therefore, we conclude that it is not feasible to
delete or alter already recorded data from OpenWPM's SQLite database.

\subsection{RQ8: Is data recording complete?}
\label{sec:incomplete_data_recording}
We investigated whether data recording is complete. We found two different
attacks against completeness: existence of unobserved channels, and silent
delivery of JavaScript code.

\noindent
\textbf{Existence of unobserved channels:} 
During our evaluation, we found a way to bypass OpenWPM's recording of
JavaScript function calls. This attack again exploits OpenWPM's hooks to
record function calls. In particular, the hooks must be attached to every
object that is to be observed. For every new window or iframe, this must
be done afresh. However, there is  a long-standing bug in Chrome and
Firefox (cf., \cite{STK17}), where both browsers under some circumstances
fail to inject scripts into iframes. We tested if OpenWPM's implementation
is affect by this and we found that this is indeed the case. 

Our evaluation of this attack involves two different ways to 
access an iframe's DOM\footnote{\texttt{window.frames[0]}, and 
\texttt{frame.contentWindow}} to create/execute iframes and their code:
static vs.~dynamic creation and immediate vs.~delayed execution. Of these,
immediate code execution (at creation time) is required to successfully
exploit this bug. None of the other parameters we tested influenced the
result. Listing~\ref{lst:silent_execution} shows a proof-of-concept of this 
type of attack.

\noindent
\textbf{Silent delivery of JavaScript code:}
Note that the aforementioned attacks based on JavaScript would appear in
OpenWPM recordings, if the HTTP instrument is used. Namely, that instrument
collects response bodies. That is, unless this instrument's recording can
also be bypassed. We indeed managed to extend our previous attacks to be
silently transferred to OpenWPM. For that, we looked at the two options that
OpenWPM offers two to collect response bodies. OpenWPM either stores all
response bodies (full coverage), or it can be set to store JavaScript files
only. The latter option significantly reduces stored content. For this mode,
we found that an attacker can silently deliver JavaScript code by sending it
as text and processing it client-side, e.g., by including a line like
\texttt{$<$link src="server.com/payload" content-type="text/plain"$>$}
in the HTML source. To successfully bypass OpenWPM's traffic recording of
JS files, three aspects must be accounted for:
\begin{enumerate}[i.,itemsep=-1ex,topsep=0ex]
\item The content-type attribute must be set to something other than
text/javascript;
\item The src attribute must not contain a ``.js''  extension;
\item the delivered file is not automatically executed; this must be
handled by a different client-side script (e.g., using \texttt{eval()}).
\end{enumerate}

\begin{lstlisting}[frame=tb, caption=Example of an unobserved channel, 
	language=js,label=lst:silent_execution, columns=fullflexible,
	basicstyle=\scriptsize,float=tbh]
// Operation will not appear in the recordings.
setTimeout(() => {
	let element = document.querySelector("#unobserved");
	let iframe = document.createElement('iframe');
	// HTML code for instantiating an iFrame 
	iframe.src = "unobserved-iframe.html"; 
	element.appendChild(iframe);
	iframe.contentWindow.navigator.userAgent;
}, 500);
\end{lstlisting}

\section{Improving OpenWPM's reliability}
\label{sec:hardening_openwpm}
\newcommand{\user}{\emph{FF}}
\newcommand{\visible}{\emph{WPM}}
\newcommand{\stealth}{\emph{WPM}$_{\mi{hide}}$}
This section focuses on OpenWPM's reliability as an instrument measuring
the web as encountered by regular visitors. We explore how and to what
extent reliability can be improved. To do so, we design an approach to
hardening OpenWPM's instrumentation and to hiding its distinctive
fingerprint (from here on referred to as \stealth). Our proof-of-concept
successfully hides the telltale signs of OpenWPM from its fingerprint and
makes OpenWPM robust in the face of the discussed attacks in a lab
setting. To evaluate its effectiveness in an open world setting, we run
\stealth{} against detectors in the wild and contrast its measurements
with those of a regular OpenWPM client.

\subsection{RQ9: How to hide the fingerprint surface?}
\label{sec:hiding}
OpenWPM's characteristic fingerprint varies with the various modes of
running OpenWPM. For example, in headless Firefox mode, the fingerprint
surface is difficult to hide due to headless mode's lack of functionality
when compared to regular browsers. Hence, we focus on run modes where
OpenWPM runs the browsers natively (Regular Mode). For such modes, we
achieve stealth by overriding properties without leaving traces. These
techniques can also be applied in other run modes (e.g., virtualisation).

The identifying properties for Regular Mode (see
Table~\ref{tbl:fp_surface}) relate to the \webdriver{} property, window
position, and dimension. Of OpenWPM's various instruments,
only the JavaScript instrument causes further identifiable properties.
Hiding these properties can be achieved by a customized browser, or by
including additional code inside a page's scope. Implementing the former
requires significant work, but it can hide the fingerprint near-perfectly.
The latter approach is far simpler to implement but risks leaving
residual traces. For our proof-of-concept, we choose the second option,
as it can be seamlessly integrated within the current OpenWPM framework
without significant effort.

Our proof-of-concept must address two aspects: hiding the automation components
and preventing detection of instrumentation. To prevent detection of instrumentation,
four issues need fixing (Sec.~\ref{sec:detecting_instrumentation}): (1) calling
the \texttt{toString} operation of overwritten functions must return the regular
output string for browser functions; (2) no additional property may appear in the
DOM; (3) stack traces must not show any signs of the instrumentation; (4) prototype
pollution must be avoided. Lastly, hiding instrumentation requires hiding their
detectable aspects, similar to how \texttt{toString} must appear unchanged.

\paragraph{Preserve \texttt{toString} output.}
For the first issue, we found that CanvasBlocker\footnote{\url{https://github.com/kkapsner/CanvasBlocker}}
addresses this well. Its implementation successfully fools all our
fingerprinting tests (Sec.~\ref{sec:identify_openwpm}).
CanvasBlocker creates a getter function with an identical signature to
the function that must be overwritten and attaches it to the DOM based on
a specific Firefox feature called \texttt{exportFunction}. The newly
exported function is then used to redefine the getter of a object's
prototype for a specific property. As a result, the overwritten function
returns the native code string like a default browser property
(cf.,~Listing~\ref{lst:openwpm_javascript}). 
Normally, accessing the getter of an object's prototype leads to an
error. If this getter is replaced with a custom getter, that error
is never thrown. This makes tampering with properties via an object's
prototype detectable~\cite{GJKKR21}. Calling the original getter
from the customised getter results in the original error being thrown,
addressing this aspect of the fingerprint surface.

\paragraph{Preserve clean DOM.}
The second issue arises during page load, prior to the page's JavaScript
activation. The instrumentation injects its code as script from
the content context into the page context, overwrites the needed
properties, and removes its code from the page context again. However,
in practice, not all injected functions are deleted. We update the
instrument to overwrite all functionality directly from the content
context, %
thus keeping the page context clean.

\paragraph{Faking stack traces.}
The third issue requires the stack trace to show no signs of instrumented
functions. A web page can only access stack traces if errors occur.
Normally, if an error occurs, the stack trace would show that the called
function is called from inside the instrumentation. We address this by
catching each error %
and throwing a new error with properly adjusted values for file name,
column, message, and line number. 

\paragraph{Avoid prototype pollution.}
The last issue relates to the pollution of an object's prototype, as
OpenWPM's instrument modifies only the first prototype in the prototype
chain. We address this by overwriting properties per prototype.
Unfortunately, this approach has its own limitation, as it is not
possible to determine the caller of a function, when a prototype has
multiple children. Especially for prototypes located higher up the
chain, the number of children increases; raising the potential to
capture unwanted API calls on other children objects. To test our
implementation, we instrumented the same API calls as used by OpenWPM.
Luckily, most of our these APIs are provided by prototypes close to the
bottom, which allows us to cover a wide set of OpenWPM's instrumented
APIs.

\paragraph{Preventing detection of automation components.}
The automation components are detectable by window size, window position 
and the webdriver attribute. For the latter, our hidden version must set the
\texttt{navigator.webdriver} property to false like a regular Firefox browser. Since 
Firefox version 88, this flag is not 
user-settable.\footnote{\url{https://bugzilla.mozilla.org/show_bug.cgi?id=1632821}}
We override the getter function of the \texttt{navigator.webdriver}
property in the same fashion as described in the previous section. To change
OpenWPM default window settings, we introduce a settings file that makes the 
window size and position settable in OpenWPM.

\subsection{RQ10: How to mitigate recording attacks?}
\label{sec:mitigation}

\paragraph{Securing messaging from page context to background context (see Sec.~\ref{sub:turn_off_recording}, \ref{sec:fake-data-injection}).}
A key benefit from migrating to Firefox's \texttt{exportFunction}, as 
described in the previous section, is the ability to export
higher privileged browser functions into the page. Hence, we can port
functionality to the page context that is otherwise only available for content
or background scripts of a browser extension. We use this to secure our
instrumented functions, as we now can use the \texttt{browser.runtime} API
to pass messages from the page to the background context. It is crucial that
such functionality is exported to a private scope of an overwritten function
to prevent access by other scripts in the page context. This prevents the
`turn recording off' and `inject fake data' attacks, as an attacker cannot
manipulate message transmissions to the background script.

\paragraph{Improving coverage of the JavaScript instrument (see Sec.~\ref{sec:incomplete_data_recording}).}
To address the tested variants of incomplete recordings, we use CanvasBlocker's
frame protection. The basic idea is to intercept APIs used by page scripts to
modify the DOM or create a new, non-instrumented copy of the DOM.
This ensures that each modification or newly constructed DOM contains the
instrumentation. Our implementation covers five cases: window constructors,
DOM modification API, window mutations, and DOM creation via the
\texttt{document.write} API, and finally the \texttt{window.open} API.

\paragraph{Filtering of the HTTP file recorder (see Sec.~\ref{sec:incomplete_data_recording}).}
To the best of our knowledge, there is no known way to distinguish JavaScript
code from text that is robust against a dedicated obfuscator. Therefore, 
an active adversary should be assumed to be capable of hiding
JavaScript in a way that would accidentally be filtered out.
Since this issue only arises in the presence of active
adversaries, we recommend in such a case not to use any filtering.

\subsection{Evaluation of PoC implementation}
\label{sec:effect_in_measurements}

We developed a proof-of-concept implementation to hide the tell-tale
signs of automation and to mitigate the found attacks.
 We evaluate the impact of our proof-of-concept implementation (from
 here on, \stealth) on web measurements when encountering bot detection
 in the wild. To that end, we contrast its results with vanilla OpenWPM
 (from here on: \visible) in HTTP traffic, cookies, JavaScript
execution, and delivered JavaScript files.
We test on all sites with bot detectors (as found by dynamic analysis)
from the Tranco Top 5K (see Sec.~\ref{sec:detector_incidence}). This list
contains 1,417 sites with either first-party or third-party detectors. On
these sites, we run \visible{} and \stealth{} in parallel (OpenWPM v.0.18.0,
regular mode, HTTP, JavaScript and cookie instrument activated) and
configure each browser to idle 60 seconds on a page after loading completed.
We take steps to mitigate noise in measurements. In particular, we avoid
cross-client interferences by separating both crawlers via two
individual machines and IP addresses. Each IP address belongs to a
residential network and comes from the same municipal and internet
provider, which avoids differences caused to cloud-based IP
blocking~\cite{ITKCPB16} and geo-location. 
Secondly, we re-synchronise the machines every 100 visit. This ensures
that sites are loaded roughly simultaneously on both machines (max.~offset
is below four minutes).

\paragraph{Sites that detect OpenWPM serve less media resources.}
\label{sec:effect_analysis}
In our experiment, we found that \stealth{} encounters 3.45\% more HTTP requests.
As our data set is not normally distributed, we tested for significance using
Wilcoxon signed-rank test with a confidence interval of 95\%. For that, we
divided the traffic into first and third-party requests and find significant
differences for HTTP requests to both first- and third-parties ($p$-value $<$
\empirical{0.0001}). In more detail, we found for \visible, 175 sites
(12\%) lead to more first-party and 472 sites (33\%) to more third-party
requests. For \stealth{}, we count 400 sites (28\%) with more
first-party and 654 sites (46\%) with more third-party requests. This
indicates a stronger variability in third-party traffic, leaning
towards less detectable clients.

Table~\ref{tbl:http_resource_types} shows requests for each machine per
requested resource type.\footnote{\url{https://developer.mozilla.org/en-US/docs/Mozilla/Add-ons/WebExtensions/API/webRequest/ResourceType}}
The table shows that \stealth{} receives roughly double the number of audio and video files
(type \emph{media}). Moreover, requested images (\emph{image} and \emph{imageset})
is increased by $\sim$3\%, and executable code (\emph{script}) by $\sim$4\%.
Moreover, \visible{} incurs three times the number of CSP violations -- though this may also be due to embedding more JavaScript in the page context.
Finally, the difference in websocket requests is due to a single outlier. Thus,
we do not expect websocket requests to change significantly between
\visible{} and \stealth{}.

\begin{table}[t]
  \centering
  \footnotesize
  \caption{Comparison of HTTP request resource types}
    \begin{tabular}{lrrr}
    \toprule
    \textbf{Resource type} & \multicolumn{1}{c}{\textbf{\visible{}}} &\textbf{\stealth{}} &
    \multicolumn{1}{c}{\textbf{Diff.}} \\
    \midrule
    csp\_report  & 884   & 298   & -66.29\% \\
    websocket  & 467   & 242   & -48.18\% \\
    media  & 378   & 552   & +46.03\% \\
    beacon  & 3,804 & 4,453 & +17.06\% \\
    imageset  & 4,888 & 5,432 & +11.13\% \\
    xmlhttprequest  & 46,199 & 49,398 & +6.92\% \\
    script  & 73,527 & 76,430 & +3.95\% \\
    object  & 53    & 55    & +3.77\% \\
    other  & 92    & 95    & +3.26\% \\
    main\_frame  & 3,883 & 3,757 & -3.24\% \\
    image  & 101,256 & 103,801 & +2.51\% \\
    sub\_frame  & 11,119 & 10,885 & -2.10\% \\
    stylesheet  & 9,663 & 9,840 & +1.83\% \\
    font  & 9,557 & 9,704 & +1.54\% \\
    \midrule
    \textbf{Total} & \textbf{265,770} & \textbf{274,942} & \textbf{+3.45\%} \\
    \bottomrule
    \end{tabular}%
  \label{tbl:http_resource_types}%
\end{table}%

\paragraph{Equivalent amount of ads/trackers traffic.}
To assess the amount of trackers and advertisers in traffic, we use the
same approach as previous works~\cite{ADZVN20,JSSBPVLK21,CLBW22}: use
the EasyList and EasyPrivacy blocklists\footnote{https://easylist.to/}
to identify trackers. Our results show that \visible{} and \stealth{}
encounter a near equal rate of advertisers and trackers. For \visible,
ads and trackers account for 14.3\% and 11.6\% of total traffic. For
\stealth{}, this is 14.2\% and 11.5\%, respectively -- almost equivalent. 

\paragraph{Large differences in served cookies.}
For cookies, we contrasted the number of cookies between both variants.
We found that these differ significantly for both first parties and third
parties ($p$-value $<$ \empirical{0.0001}). Specifically, 305 sites serve
\stealth{} more first-party cookies, while only 146 sites serve \visible{} more
first-party cookies. Interestingly, the opposite is true for third-parties.
Here we find 824 sites whose third parties offer \visible{} more cookies than
\stealth{}; the other way around happens for the third parties of 227 sites.
In total, the number of cookies is 55,853 (\visible{}) vs.~46,736 
(\stealth{}). Using \stealth{} thus leads to a decrease of 16.32\% of cookies.

We also looked at cookies as possible means to track users. To determine
whether a cookie can be used for web tracking, we use the approach of Englehardt
et al.~\cite{EREZ*15}, as refined by Chen et al.~\cite{CIPK21}. According to
this method, a cookie may be used for tracking when: (1) it cannot be a
session cookie, (2) the length of the cookie is 8 or more characters (excluding
surrounding quotes), (3) the cookie is always set, and (4) the values differ
significantly based on the Ratcliff-Obershelp algorithm~\cite{Ratcliff}.
While 5,307 cookies satisfy these criteria for \visible{}, only 2,282 cookies for \stealth{}
match; a decrease of 57\%.

\hide{
Measurement studies investigating tracking/advertising typically use blocklists
to gauge advertising and tracking. We follow suit, using the
Easylist and EasyPrivacy blocklists. react differently to \stealth{}.
We do not claim these lists as complete in any sense, but rather as rough
indicator.

Table~\ref{tbl:traffic_overview} shows that \empirical{19\%} of all requests are
tagged by these list as known advertisers and trackers. Despite the increased
volume of requests in \stealth{}, the proportion of ad and tracker traffic is
small; we see an increase of 0.8\% for EasyList and 1\% for EasyPrivacy in \stealth{}
traffic. We find that less than half of the tracking domains treat both bots equally;
specifically, EasyList: \empirical{274} of \empirical{570} and EasyPrivacy:
\empirical{122} of \empirical{271}. Figure~\ref{fig:tp_el_ep} breaks down
non-zero differences in number of requests between \visible{} and \stealth, for
domains in either list.
We would have expected the requests to strictly increase when hiding the 
fingerprint
(as some trackers would stop tracking once they detected a bot).
Interestingly, we find that there are tracker/advertiser domains that are more 
prevalent
for \visible{} (incl.~pubmatic.com and gstatic.com), while others occur more often 
for
\stealth{} (incl.~doubleclick.com and google-analytics.com).
}

\hide{
\paragraph{JavaScript execution.}
\todo{
\begin{itemize}
    \item How many of these scripts execute fingerprinting routines (e.g. webdriver and userAgent) in iframes that where not captured by the original extension
    \item How often was \stealth{} detected compared \visible{}
    \item Port scanning? A bit boring, as this was only one site
\end{itemize}

}
}

\section{Conclusions}

\paragraph{Reliability of automated measurements on trial.} 
Our work demonstrates that OpenWPM is susceptible to attacks threatening
its reliability. 
In particular: virtualisation makes scaling web
studies easy, but turned out to undermine OpenWPM's reliability as
a measurement tool. It is an open question whether other automation /
measurement frameworks suffer similarly from virtualisation.

\paragraph{Bot detection on the rise.}
In comparison with previous studies, we see the number of sites looking
for the \webdriver{} property has significantly increased in the span of
less than one year (Tbl.~\ref{tbl:related_work_bot_detectors}).
This rapid change clearly suggests that web sites are swiftly transitioning
to responding differently to automated clients than to regular clients. Web
studies should therefore no longer ignore the potential impact of bot detection
on their study.

\begin{table}[t]
\centering
\footnotesize
\caption{Studies measuring \webdriver{} property access on front pages}
\begin{tabular}{llllrr}
	\toprule
	& \textbf{when} & \textbf{analysis} &  \textbf{corpus} &
	\textbf{\# sites} & \textbf{\%}  \\
	\midrule
	\cite{JK19} & 2019--10 & dynamic  & Alexa 50K   &  2,756  &  5.51\% \\
	This paper   & 2020--07 & combined & Tranco 100K & 13,989 & 13.99\% \\
	&          & -- \emph{static}  &    & \emph{11,957}     & \emph{11,96}\% \\
	&          & -- \emph{dynamic} &    & \emph{12,194}     & \emph{12.19}\% \\
	
	\bottomrule
\end{tabular}
\label{tbl:related_work_bot_detectors}
\end{table}

\paragraph{Towards robust instrumentation.}
\label{sec:alternative-way-of-fixing}
Our findings highlight the difficulties of deploying instruments
via the page context. %
To improve robustness, we
advocate moving the instruments outside of page scope.
To achieve this, the debugger API
could be leveraged. However, OpenWPM uses Selenium v3, which does not
support this (planned for Selenium v4). Alternatively, instrumentation
could be integrated in the browser's source code. This would give great
flexibility in hiding distinctive aspects of the browser fingerprint.
This would also incur significant additional maintenance overhead
slowing adoption of new browser versions. However, OpenWPM's rate of
adoption is already slow -- the tradeoff may thus be worth it.

\paragraph{Advice for conducting a web measurement study.} 
While the evaluation of our proof-of-concept is limited in scope, we
still find significant differences in a variety of attributes. While
studies that focus on the amount of traffic seem to be (for now) in the
clear, studies that focus on audio/video files or web tracking via
cookies must take bot detection into account (Table~\ref{tbl:http_resource_types}).
Similarly, studies that automatically crawl beyond the front page will
encounter more bot detectors (Table~\ref{tbl:literature_review}).

\paragraph{Ethics.}
Our work aims to make OpenWPM a more reliable measurement framework. 
We responsible disclosed our findings and shared fixes of the identified issues. 
This helps make OpenWPM less detectable, and therefore its results more reliable.
Of course, a less detectable scraper may itself be abused. 
For attacking specific sites, our improvements do not greatly impact the attack
surface: a less detectable OpenWPM is a fine tool for studying thousands of
sites, but not for a targeted attack on a specific site.
For attacks that span thousands of sites (e.g., clickfarming), our improvements
do not help:  disguising as a regular browser is insufficient to overcome
contemporary defenses.
For that, site-specific fingerprints are needed~\cite{TJM15}. Thus, existing
re-identification-based countermeasures (e.g., rate limiting) are not impacted.

\paragraph{Availability \& responsible disclosure.} 
Our stealth extension is available via 
GitHub.\footnote{\url{https://github.com/bkrumnow/OpenWPM/tree/stealth\_extension}}
We disclosed our findings (both attacks and identifiable properties) to
the OpenWPM developers. We are working towards having our fixes integrated
into the framework.

\bibliographystyle{plain}
\bibliography{literature}

\appendix

\section{Patterns for common first-party detectors}
\label{appdx:first_party_detector_details}
Table~\ref{tbl:first_party_detector_patterns}, shows patterns we found
in our first-party script analysis from
Section~\ref{sec:detectors_analysis}. Scripts provided by Akamai,
Incapsula, Cloudflare, and PerimeterX follow the same script pattern,
these can be easily recognised. For the unknown script, we found that
the for common path patterns between larger clusters of script hashes. A
manual validation showed that scripts found under the listed path are
most similar.

\begin{table}[htbp]
	\centering
	\small
	\caption{Similarities in first-party detectors}
	\setlength{\tabcolsep}{5pt}
	\begin{tabular}{llr}
		\toprule
		\textbf{Origin} & \textbf{URL path similarities} & \textbf{\# sites}\\
		\midrule
		Akamai   & domain/akam/11/\ldots & 1,004\\
		Incapsula  & domain/\_Incapsula\_Resource?\ldots  & 998 \\
		Unknown & domain/asssets/\{hash of 31-32 bits length\}  & 659 \\
		& domain/resources/\{hash of 32-33 bits length\} & \\
		& domain/public/\{hash of 32-33 bits length\} & \\
		& domain/static/\{hash of 34 bits length\} & \\
		Cloudflare & domain/\ldots/cdn-cgi/bm/cv/2172558837/api.js& 486	\\
		PerimeterX & domain/\ldots{}/\{8 character string\}/init.js & 134 \\
		\bottomrule
	\end{tabular}
	\label{tbl:first_party_detector_patterns}
\end{table}

\hide{
\section{Reducing measurement noise}
\label{appdx:reducing_noise}

To properly attribute observed effects to bot detection, we need to
address various factors for noise. We consider four factors that we
discuss in detail below.

\paragraph{Cross-client interference}.
During the validation of \stealth, we observed that bot
detection resulted in either block pages for one machine regardless of
the used browser, or for all machines operating under the same IP
address. To prevent such effects from skewing the results of the
experiment in this section, we use two different machines with different
IP addresses, running the same OS: one for OpenWPM and one for \stealth.

\paragraph{Differences based on IP address origin.}
Previous studies have shown that the origin of IP addresses used for
automated web clients may affect the studies outcome. Specifically,
blocking~\cite{MBVVSSHE18} and deviating content~\cite{KSHL15,HTWH18}
may occur based on a client's IP-based geolocation. Further, multiple
studies~\cite{ZBORSWL20,ITKCPB16} have attributed observed differences
to cloud- vs.~university-based IP addresses. To mitigate such effects
within our experiment, we use two residential IP addresses from the same
local ISP. The two machines were run from the same municipality but
located in different buildings.

\paragraph{Differences based on page dynamics.} 
Websites inherently change over time for various reasons: A/B testing,
content updates, layout changes, targeted advertisements, etc. Such
factors make a one-on-one comparison fragile. Even if a site is loaded
in two browsers simultaneously, the resulting pages may be different. We
account for this by visiting each site 8 times from the same machine, to
establish a baseline of website-inherent changes. Specifically, we run 8
browser instances per machine, which each visit the same set of sites
simultaneously (requests are synchronised). We build one baseline for
each site per OpenWPM version (\visible{} and \stealth{}). While this
approach will significantly reduce or eliminate noise due to page
dynamics, a caveat is
that no user would request the same page 8 times simultaneously. Thus,
this approach makes us more detectable on an HTTP-level. We consider
this an acceptable trade-off in order to establish a baseline for
comparison.

While site visits are individually synchronised within one test machine,
synchronisation with the other test machine is per-batch. We start each
machine with the same limited set of sites at the same time and avoid
browser restarts as these increase desynchronisation. Throughout our
measurements we observed a maximum variation of 3 minutes between the
two machines loading the same site.

\paragraph{Validation corpus.}
To ensure the correctness of the result, we manually verify screenshots
of the scraping runs. This does limit the number of sites that we can
feasibly test. Therefore, we select a random subset of sites from the
Tranco Top 10K. We separate the 10K list into 10 buckets with 1,000
sites each. We take a random sample of 100 sites from each of these
buckets, which results into a final set of 1,000 sites.
}
\section{Adoption of new Firefox versions by OpenWPM.}
\label{appdx:fx_versions}
Releases of OpenWPM do not appear synchronously with
Firefox. As a result, certain time frames exist where the OpenWPM client
uses an older Firefox versions than regular users.
Table~\ref{tbl:firefox_releases} summarises migration of Firefox
versions in the OpenWPM Framework since version 0.10.0. Between the
release of Firefox 77 (March 2020) and the release of OpenWPM v.0.18.0,
amounts to 561 days. Within this period, OpenWPM was shipped with an outdated version
398 days (71\%). %

\begin{table}[ht]
	\centering
	\footnotesize
	\caption{Integration of Firefox releases into OpenWPM}
	\begin{tabular}{rrrrr}
		\toprule
		\textbf{Firefox} & \textbf{release date} & \textbf{OpenWPM} &  \textbf{integration date} & \textbf{Outdated}\\
		\midrule
		95.0    & 12/07/21 & 0.18.0 & 12/16/21 & 69 days\\
		94.0    & 11/02/21 &       &  \\
		93.0    & 10/05/21 &       &  \\
		92.0    & 09/07/21 &       &  \\
		91.0    & 08/10/21 &       &  \\
		90.0    & 07/13/21 & 0.17.0 & 07/24/21 & 11 days\\
		89.0    & 06/01/21 & 0.16.0 & 06/10/21 & 9 days\\
		88.0    & 04/19/21 & 0.15.0 & 05/10/21 & 48 days\\
		87.0    & 03/23/21 &       & \\
		86.0.1  & 03/11/21 & 0.14.0 & 03/12/21 & 87 days\\
		85.0    & 01/26/21 &       &  \\
		84.0    & 12/15/20 &       &  \\
		83.0    & 11/18/20 & 0.13.0 & 11/19/20 & 58 days \\
		82.0    & 10/20/20 &       & \\
		81.0    & 09/22/20 &       & \\
		80.0    & 08/25/20 & 0.12.0 & 08/26/20 & 29 days\\
		79.0    & 07/28/20 &       &  \\
		78.0.1  & 07/01/20 & 0.11.0 & 07/09/20 & 8 days\\
		77.0    & 06/03/20 & 0.10.0 & 06/23/20 & 20 days\\
		\bottomrule
	\end{tabular}
	\label{tbl:firefox_releases}%
\end{table}

\hide{
\section{Mitigation} 
\label{appx:mitigation}
To mitigate cloaking-based attacks, it is indispensable to effectively hide identifiable properties. Making OpenWPM less
detectable will pressure malicious actors to move to different
detection approaches. Other detection approaches do not work as
seamlessly as fingerprinting -- the need to collect more
interaction steps prevents immediate detection -- which will make
cloaking less threatening. 

Vulnerability 1, preventing data recording, can be mitigated by also instrumenting the setter of the event dispatcher in JavaScript. While it will not prevent the attack, the data recording will show that the event dispatcher was tampered with. When this happens, that means that either data recording was prevented, and/or fake data was injected (vulnerability 2).

The first variant of vulnerability 4, using unobserved channels, can be
mitigated by JavaScript functions that patch the underlying bug. Several Firefox extensions (e.g.~User-Agent
Switcher\footnote{\url{addons.mozilla.org/en-US/firefox/addon/uaswitcher}}
and 
CanvasBlocker\footnote{\url{addons.mozilla.org/en-US/firefox/addon/canvasblocker}}) already include functionality that (partially) addresses this. We tested these in a preliminary test run and found that some, but not all cases were fixed. 
The second variant of vulnerability 4, silently delivering JavaScript code, works by delivering JavaScript as text files. Post-delivery detection is too complex: it requires distinguishing between arbitrary text and JavaScript, and will lead to an arms race where JavaScript is hidden more and more cleverly. As such, we recommend full HTTP recording when correct recording of received JavaScript files needs to be guaranteed.
}

\hide{
\paragraph{Recommendations for future studies.}
Based on our investigation, we offer the following recommendations. 

\noindent
For measurement studies:
\begin{itemize}
	\item {Avoid using virtualisation}
	\item {Hide the \webdriver{} attribute}
	\item Do not use a headless browser
	\item Be careful using the built-in interactions 
\end{itemize}
For OpenWPM developers:
\begin{itemize}
	\item {Reduce OpenWPM's fingerprint}
	\item {Account for bot detection}. \\
	Incorporate measures to identify bot detection and mark data
	from sites with bot detection as such.
	\item Offer a new interaction API that generates less distinct
	behavioral patterns.
\end{itemize}
While the above recommendations may not fully eliminate the effects of
bot detection, they will certainly reduce it.
}

\section{Patterns used in static analysis}
\label{appdx:patterns}
We iterate on the pattern design to reduce false positives. Our very
first run used patterns matching strings literally. However, in the
specific case of matching the term \emph{webdriver}, we found that this
selects scripts that use this word in another context than checking
Selenium-driven Firefox browsers (cf.~\cite{JK19,JKV19} for conflicting
bot detection properties with this term). In the next iteration we used
patterns that take the context of the access to a property into account.
For example, the pattern
\texttt{navigator\textbackslash[["\textquotesingle]webdriver["\textquotesingle]\textbackslash]}
 only matches if the \webdriver{} property is checked via the navigator 
object. Table~\ref{tbl:patterns} lists our explored patterns. Finally, we manually 
checked a random subset to check pattern performance. Only one pattern still 
introduced false positives; all its matches were manually validated and false 
positives eliminated.

\begin{table}[ht]
	\centering
	\footnotesize
	\caption{Patterns evaluated in static analysis}
	\begin{tabular}{lr}
		\toprule
		\textbf{Pattern} & \textbf{false positives found} \\
		\midrule
		\webdriver{} & \checkmark\\
		\texttt{instrumentFingerprintingApis} & - \\
		\texttt{getInstrumentJS} & - \\
		\texttt{jsInstruments} & - \\
 	   \texttt{(?<!\_|-)webdriver(?!\_|-)} & \checkmark\\
		\texttt{navigator.webdriver} & - \\
		\texttt{navigator\textbackslash[["\textquotesingle]webdriver["\textquotesingle]\textbackslash]}
		& -\\
		\bottomrule
	\end{tabular}
	\label{tbl:patterns}%
\end{table}

\section{Previous studies relying on OpenWPM}
\label{appdx:previous_studies}
\newcommand{\xmark}{x}
\renewcommand{\xmark}{\ensuremath{\circ}}
Table~\ref{tbl:openwpm_studies_full} provides a detailed view on our 
analysis of previous peer-reviewed studies based on OpenWPM. Each category 
that applies to a study is marked with a ``\checkmark{}''. For those studies that 
measure certain aspects, but rely on out of bound mechanisms (e.g., by deploying 
a proxy) and do not rely on OpenWPM's instrumentation are marked with a 
``\xmark{}''. Running modes are shortened in the table as follow: unspecified (u), 
native (n), headless (h), xvfb (x), docker (d), virtual machine (v). Papers that are not 
included in the seed list, but where added by us, are highlighted with a 
``$\star$''.

\begin{table*}[htbp]
\centering
\footnotesize
\caption{Overview of previous studies using OpenWPM for web studies}
\begin{tabular}{lrr cc ccc ccc c cc}
    \toprule
    &&& \multicolumn{2}{c}{deployed as}& \multicolumn{3}{c}{measures} & \multicolumn{3}{c}{uses}& visits & uses & mentions
    \\
    \cmidrule(lr){4-5} \cmidrule(lr){6-8} \cmidrule(lr){9-11} \cmidrule(lr){12-12} \cmidrule(lr){13-13} \cmidrule(lr){14-14}
    \textbf{Year} & \textbf{Ref.} & \textbf{1$^{st}$ Author} & \textbf{Mode} & \textbf{VM} & 
    \textbf{Cookies} & \textbf{HTTP}  & 
    \textbf{JS} & \textbf{Scrolling} & \textbf{Clicking} & \textbf{Typing} & 
    \textbf{Sub-pages} & \textbf{Anti-BD} & \textbf{BD} 
    \\
    
    \cmidrule(lr){1-4}
    \cmidrule(lr){4-5} \cmidrule(lr){6-8} \cmidrule(lr){9-11} \cmidrule(lr){12-12} \cmidrule(lr){13-13} \cmidrule(lr){14-14}
    2014 & \cite{AEEJ*14} &  Acar & u & \checkmark &\xmark{} & \xmark{} & \checkmark &&&&&&\\
    & \cite{RB14} &  Robinson & u && &&&& \checkmark & \checkmark &&&  \\
    
    \cmidrule(lr){1-3}
    \cmidrule(lr){4-5} \cmidrule(lr){6-8} \cmidrule(lr){9-11} \cmidrule(lr){12-12} \cmidrule(lr){13-13} \cmidrule(lr){14-14}
    2015 & \cite{EREZ*15} &  Englehardt & u & \checkmark & \checkmark & \checkmark &&&&&&&\\
    & \cite{KB15} &  Kranch & u && \checkmark & \xmark{} &&&&&&& \\
    & \cite{AGH15} &  Altaweel & h &&\checkmark & \checkmark &&& \checkmark && \checkmark &&  \\
    & \cite{FMSB15} &  Fruchter & u & \checkmark &\checkmark & \checkmark &&&&&&&\\
    \cmidrule(lr){1-3}
    \cmidrule(lr){4-5} \cmidrule(lr){6-8} \cmidrule(lr){9-11} \cmidrule(lr){12-12} \cmidrule(lr){13-13} \cmidrule(lr){14-14}
    2016  & \cite{AJ16} &  Andersdotter & u && \checkmark & \checkmark &&&&&\checkmark&&  \\
    & \cite{EN16} &  Englehardt & x & \checkmark & \checkmark & \checkmark & \checkmark &&&&\checkmark&&  \\
    & \cite{SDAHN16} &  Starov & u & \checkmark && \checkmark &&&&&&&\\
    \cmidrule(lr){1-3}
    \cmidrule(lr){4-5} \cmidrule(lr){6-8} \cmidrule(lr){9-11} \cmidrule(lr){12-12} \cmidrule(lr){13-13} \cmidrule(lr){14-14}
    2017  & \cite{MSN17} &  Miramirkhani & u & \checkmark && \xmark{} & \checkmark &&\checkmark&&&&  \\
    & \cite{BRAY17} &  Brookman & u & \checkmark & \checkmark & \checkmark &&&&&&&  \\
    & \cite{RK17} &  Reed & u &&& \checkmark &&&&&&&  \\
    & \cite{OEN17} &  Olejnik & u &&&& \checkmark &&&&&&  \\
    & \cite{MWPH17} &  Maass & u && \checkmark & \checkmark &&&&&&&  \\
    & \cite{LWPY*17} &  Liu & h &&&&&&&&&&  \\
    & \cite{Schmeiser17} &  Schmeiser & u &&& \checkmark &&&&&&&  \\
    
    \cmidrule(lr){1-3}
    \cmidrule(lr){4-5} \cmidrule(lr){6-8} \cmidrule(lr){9-11} \cmidrule(lr){12-12} \cmidrule(lr){13-13} \cmidrule(lr){14-14}
    2018  & \cite{GKRN18} &  Goldfeder & u &&& \checkmark &&&&&&&  \\
    & \cite{EHN18} &  Englehardt & u &&& \checkmark &&& \checkmark & \checkmark & \checkmark && \checkmark \\
    & \cite{BZKS18} &  Binns & h && \checkmark & \checkmark &&&&&&&  \\
    & \cite{DABP18} &  Das & u &&& \checkmark & \checkmark &&&&&& \checkmark \\
    & \cite{AHS18} & van Acker & u &&& \checkmark &&&&&&&  \\
    & \cite{DMF18} &  Dao & u &&& \checkmark &&&&&&&  \\
    
    \cmidrule(lr){1-3}
    \cmidrule(lr){4-5} \cmidrule(lr){6-8} \cmidrule(lr){9-11} \cmidrule(lr){12-12} \cmidrule(lr){13-13} \cmidrule(lr){14-14}
    2019  & \cite{CHPN19} &  Cozza & u &&&&& \checkmark & \checkmark & 
    \checkmark & \checkmark &&  \\
    & \cite{GZDM19} &  Gomes & u &&& \checkmark &&&&&&&  \\
    & \cite{EAWN19} &  van Eijk & d &&&&&&&&&&\\
    & \cite{SK19} & Sørensen & u & \checkmark && \checkmark &&&&& \checkmark &&  \\
    & \cite{LDAW19} &  Liu & u &&& \checkmark &&&&&&& \checkmark \\
    & \cite{MAFL19} &  Mathur & u &&& \checkmark &&&\checkmark&&\checkmark&&  \\
    & \cite{RUW19} &  Ramadorai & u &&\checkmark& \checkmark &&&&&&&  \\
    & \cite{MGF19} &  Mazel & u && & \checkmark &&&&&&  \\
    & \cite{AOMY19} &  Ali & u && \checkmark &&&&&&&&  \\
    & \cite{SM19a} &  Samarasinghe & u && \checkmark & \checkmark &&&&&&& \checkmark \\
    & \cite{MSH19} &  Maass & u &&& \checkmark &&&&&&&  \\
    & \cite{SIIK19} &  Solomos & u &&&&& \checkmark & \checkmark &&&&  \\
    & \cite{VFGVA19} &  Vallina & u && \checkmark & \checkmark & \checkmark &&&&&&  \\
    & \cite{JKV19} &  Jonker & h && \checkmark && \xmark{} &&&&&& 
    \checkmark \\
    & \cite{UTDHP19} &  Urban & u && \checkmark & \checkmark &&&&& \checkmark &&  \\
    & \cite{SM19b} &  Sakamoto & u && \checkmark &&&&&&&&  \\
    
    \cmidrule(lr){1-3}
    \cmidrule(lr){4-5} \cmidrule(lr){6-8} \cmidrule(lr){9-11} \cmidrule(lr){12-12} \cmidrule(lr){13-13} \cmidrule(lr){14-14}
    2020  & \cite{FBLS20} &  Fouad & u && \checkmark & \checkmark &&&&& 
    \checkmark &&  \\
    & \cite{CNS20} &  Cook & u &&&&& \checkmark &&&& \checkmark & \checkmark \\
    & \cite{YY20} &  Yang & u && \checkmark & \checkmark & \checkmark &\checkmark&&&&& \\
    & \cite{AEN20} &  Acar & u & \checkmark && \checkmark& \checkmark &&&& \checkmark &\checkmark & \checkmark\\
    & \cite{KTK20} &  Koop & d && \checkmark & \checkmark & \checkmark && \checkmark &&& \checkmark & \\
    & \cite{ZBORSWL20} &  Zeber & n/x & \checkmark & \checkmark &\checkmark & \checkmark &&&&&& 
    \checkmark \\
    & \cite{ADZVN20} &  Ahmad & u &&& \checkmark &&&&&& \checkmark & \checkmark \\
    & \cite{AJPSK20} &  Agarwal & h & \checkmark & \checkmark & \checkmark & \checkmark &&&&&&  \\
    & \cite{UDHP20} &  Urban & u && \checkmark & \checkmark & \checkmark &\checkmark &&& \checkmark & \checkmark & \checkmark \\
    & \cite{UTDHP20} &  Urban & u & \checkmark & \checkmark & \checkmark && \checkmark 
    &&& \checkmark & \checkmark & \checkmark \\
    & \cite{PDAGN20} & Pouryousef & u && \checkmark  &&&&&&&&  \\
    & \cite{FSKBC20} &  Fouad & u && \checkmark & \checkmark &&&&&&&  \\
    & \cite{SCLN20} & Sivan-Sevilla & u & \checkmark & \checkmark & \checkmark 
    & \checkmark &&&&& \checkmark & \checkmark \\
    & \cite{HTS20} &  Hu & u && \checkmark & \checkmark &&&&&&&  \\
    & \cite{DF20a} &  Dao & u &&& \checkmark &&&&&&&  \\
    & \cite{SIK20} &  Solomos & n && \checkmark & \checkmark &&&&&&&  \\
    & \cite{DF20b} &  Dao & u &&& \checkmark &&&&&&&  \\
    
    \cmidrule(lr){1-3}
    \cmidrule(lr){4-5} \cmidrule(lr){6-8} \cmidrule(lr){9-11} \cmidrule(lr){12-12} \cmidrule(lr){13-13} \cmidrule(lr){14-14}
    2021  & \cite{CUTSS21} &  Calzavara & u && \checkmark &
    \checkmark &&&&& \checkmark && \checkmark \\
    & \cite{RTM21} &  Rizzo & u & \checkmark && \checkmark &
    \checkmark &&&&\checkmark&&  \\
    & \cite{IES21} &  Iqbal & u &&& \checkmark & \checkmark&&&&&&\\
    & \cite{GJKKR21} & Go{\ss}en$^{\star}$ & n &&& \checkmark &&\checkmark&\checkmark&\checkmark&&&\checkmark\\
    \cmidrule(lr){1-3}
    \cmidrule(lr){4-5} \cmidrule(lr){6-8} \cmidrule(lr){9-11} \cmidrule(lr){12-12} \cmidrule(lr){13-13} \cmidrule(lr){14-14}
	2022 & \cite{CLBW22} &  Cassel$^{\star}$ & u 
    && &\xmark&\xmark &&&&&&\checkmark\\
    \bottomrule
\end{tabular}%
\label{tbl:openwpm_studies_full}%
\end{table*}%

\end{document}